\documentclass[10pt,notitlepage,nofootinbib,preprintnumbers, aps,prd,superscriptaddress,raggedbottom]{revtex4-1}
\usepackage{amsmath,amssymb,natbib,bm,color,xcolor}
\usepackage{appendix}
\usepackage{graphicx}
\usepackage{comment}

\usepackage[colorlinks=true,citecolor=black,urlcolor=black,linkcolor=black]{hyperref}

\def\tev{\, {\rm TeV}}
\def\gev{\, {\rm GeV}}

\begin{document}
\title{WIMPs with Enhanced Annihilation from a Feebly Interacting Unstable Partner}
\author{Chance Hoskinson}
\email{chance.hoskinson@utah.edu}
\affiliation{Department of Physics and Astronomy, University of Utah, Salt Lake City, UT, 84112, USA}
\author{Pearl Sandick}
\email{pearl.sandick@utah.edu}
\affiliation{Department of Physics and Astronomy, University of Utah, Salt Lake City, UT, 84112, USA}
\author{Barmak Shams Es Haghi}
\email{barmak.shams@utah.edu}
\affiliation{Department of Physics and Astronomy, University of Utah, Salt Lake City, UT, 84112, USA}
\affiliation{Texas Center for Cosmology and Astroparticle Physics, Weinberg Institute for Theoretical Physics, Department of Physics, The University of Texas at Austin, Austin, TX, 78712, USA}

\begin{abstract}  
Weakly interacting massive particles (WIMPs) constitute one of the best-motivated dark matter (DM) candidates. In this study, we consider a simple extension of the WIMP paradigm in which WIMPs are coupled to a heavier particle that interacts only with them and can decay into WIMPs. In the early Universe, when WIMPs are in thermal equilibrium with the Standard Model (SM) bath, they serve as a portal to populate the heavy partner through inverse decays. Depending on the lifetime of the partner particle, its eventual decay into WIMPs can significantly alter the WIMP abundance. If the partner decays into WIMPs before their freeze-out from equilibrium, the standard thermal history remains unchanged. When the decay occurs after WIMP freeze-out, however, the injected non-thermal population of WIMPs can result in an overproduction of DM, therefore requiring a larger WIMP annihilation cross section relative to the standard scenario without the partner. For a given partner lifetime and WIMP annihilation cross section, freeze-out may be delayed due to re-annihilation or followed by a freeze-in phase. By solving the Boltzmann equations both semi-analytically and numerically, we explore these thermal histories and study the required enhancement in the WIMP annihilation cross section compared to the canonical value. This enhancement, which can reach up to three orders of magnitude, leads to stronger annihilation signals in the late Universe. As a result, current and upcoming indirect detection experiments are able to probe the lifetime of the partner particle. We briefly explore possible connections between our scenario and the baryon asymmetry of the Universe. As an example, we show that the Next-to-Minimal Supersymmetric Standard Model (NMSSM) can provide a realization of our scenario, with the possibility of accommodating light higgsino DM with the correct relic abundance.  
\end{abstract}
 \maketitle
\section{Introduction}
Weakly interacting massive particles (WIMPs) have long been considered a leading candidate for dark matter (DM). As a thermal relic, the final abundance of WIMPs is set by their freeze-out from the Standard Model (SM) thermal bath in the early Universe; as the temperature drops below the mass of the WIMP, its production from the thermal bath becomes exponentially suppressed. Eventually, the Hubble expansion rate exceeds the annihilation rate, and the WIMPs drop out of thermal equilibrium. Assuming that WIMPs are cold relics, as required for consistency with structure formation in the Universe, together with unitarity limits on WIMP interactions with SM particles, points to a GeV-TeV mass range. Furthermore, reproducing the observed dark matter abundance today requires a weak-scale WIMP annihilation cross section. More specifically, for a Majorana fermion WIMP the observed relic abundance is obtained for the canonical annihilation cross section $\langle\sigma v\rangle_{\rm th}=2.2\times 10^{-26}\,{\rm cm^3}\,{\rm s^{-1}}$~\cite{Steigman:2012nb}. As a result, WIMPs are a well-motivated target for a broad search program, including collider searches~\cite{Boveia:2018yeb},  direct detection experiments~\cite{LZ:2024zvo,XENON:2025vwd} and indirect detection searches~\cite{Fermi-LAT:2016uux}.
 
In this paper, we investigate a simple scenario that extends the WIMP paradigm by introducing a new unstable partner, referred to as the WIMP partner, that interacts exclusively with the WIMPs. In this setup, WIMPs, which in the early Universe are a part of the thermal bath, act like a portal to produce their partners. We assume that the WIMP partner is at least twice as heavy as the WIMPs and eventually decays into them. As we show in detail, the lifetime of the partner leads to three distinct thermal histories and mechanisms for setting the relic abundance of the WIMPs: regular freeze-out from equilibrium abundance, freeze-out from re-annihilation, and regular freeze-out from equilibrium abundance followed by freeze-in. 

If the coupling between the WIMP and its partner particle is large enough that the partner decays while the WIMP is still in chemical equilibrium with the thermal bath, i.e., before WIMP freeze-out from equilibrium abundance, the WIMPs produced in these decays rapidly attain chemical equilibrium with the bath. Consequently, the WIMP thermal history, its freeze-out, and final relic abundance remain unchanged.

The partner affects the WIMP thermal history only if its coupling to the WIMP is sufficiently small for it to decay around the time of WIMP freeze-out from equilibrium abundance or later. If the coupling is sufficiently weak, the partner can survive long enough to decay into WIMPs after the standard WIMP freeze-out from its equilibrium abundance, which typically occurs when the bath temperature drops to approximately $\sim 0.05$ times the WIMP mass~\cite{Kolb:1990vq}. In this case, a non-thermal population of WIMPs is injected into the thermal bath. Depending on the lifetime of the partner, these newly produced WIMPs may undergo an additional period of efficient annihilation known as re-annihilation. During re-annihilation, the WIMP yield tracks a quasi-equilibrium curve, then departs from it and freezes out to its final abundance. As we show, because freeze-out from re-annihilation occurs later than the standard freeze-out from equilibrium, the re-annihilation phase results in a larger final WIMP yield than conventional freeze-out. Consequently, reproducing the observed relic abundance today requires an annihilation cross section larger than the canonical value. This simple setup therefore allows WIMPs to possess an enhanced annihilation cross section and yield the correct relic abundance today. The enhancement of the annihilation cross section, which can be as large as three orders of magnitude, is inversely proportional to the feeble coupling between the WIMP and its partner particle. For smaller couplings, the partner particle decays at later times. Eventually, the decay occurs only after the WIMP number density remaining after thermal freeze-out from equilibrium has become highly diluted. At this stage, the injected WIMPs can no longer annihilate efficiently to track the quasi-equilibrium value. Consequently, re-annihilation is no longer effective, and the injected WIMPs simply add to the frozen-out abundance. This can be interpreted as a freeze-in production channel for WIMPs~\cite{Hall:2009bx}, in which the final DM abundance is determined by freeze-out from equilibrium abundance followed by freeze-in. In this case as well, the WIMP annihilation cross section must exceed its canonical value to suppress the freeze-out contribution and leave room for the subsequent freeze-in contribution to reproduce the observed relic abundance. In this case, the required enhancement is generally not significant except within a narrow range of coupling values. 

Depending on the mass of the WIMP and its annihilation products, annihilation cross sections exceeding the canonical thermal value by up to three orders of magnitude are allowed and can be probed by current and upcoming indirect detection experiments. As a result, our scenario can be tested through these searches, which, as we show, in turn probe the lifetime of the feebly interacting WIMP partner. 

Beyond its prospects for indirect detection, a particularly interesting realization of our framework arises within the NMSSM, where light (sub-TeV) higgsino DM with the correct relic abundance can be realized.
 
Enhancing the WIMP annihilation cross section beyond its canonical value while simultaneously reproducing the observed relic abundance is a nontrivial task. Here, we achieve this within a simple and minimal framework, without invoking non-standard cosmologies. A similar result can be obtained, for instance, in scenarios where a non-thermal population of WIMPs is produced through gravitational interactions in addition to the thermal population from the radiation bath~\cite{Gondolo:2020uqv}. Other examples include scenarios where WIMP production and freeze-out occur during an early matter-dominated era~\cite{Kawasaki:1995cy,Moroi:1999zb,Acharya:2009zt,Choi:2018kto}, or where a first-order phase transition modifies the WIMP mass during its thermal evolution~\cite{Allahverdi:2024ofe}.

This paper is organized as follows. In Section~\ref{sec:model}, we present our simple model. Then, we explore three different thermal histories of WIMP DM that can arise from decay of WIMP partner in Section~\ref{sec:history}. In Section~\ref{sec:constraints}, we discuss the constraints on our model from indirect detection searches and cosmological observations. Section~\ref{sec:baryogenesis} provides a brief discussion that explains the possible connections between our model and baryon asymmetry of the Universe. In Section~\ref{sec:NMSSM}, we explore the possibility of realization of our model within the NMSSM and its implications for DM candidates in this motivated particle physics model. We conclude in Section~\ref{sec:conclusion} by summarizing our minimal model and its implications, and outlining possible future directions.
\section{Model}
\label{sec:model}
We assume that a Majorana fermion DM candidate, $\chi$, has interactions with the SM that keep it in thermal equilibrium with the thermal bath in the early Universe. The mass of $\chi$ is assumed to be in the range GeV-TeV, and therefore it is a WIMP DM candidate. As noted above, for such a WIMP, a
thermally averaged annihilation cross section of the order of the canonical annihilation cross section, $\langle\sigma v\rangle_{\rm th}$, leads to the freeze-out of WIMPs with the observed relic abundance today. We also assume that there is a new real scalar in the spectrum, $\phi$, which only interacts with the DM through a feeble interaction with Yukawa coupling $\lambda$:
\begin{equation}
    \mathcal{L}\supset -\frac{1}{2}m^2_\phi\phi^2-\left(\frac{1}{2}m_\chi\chi\chi+\frac{1}{2}\lambda\phi\chi\chi+{\rm h.c.}\right),
\end{equation}
and we require $m_\phi>2m_\chi$.  
\section{Thermal Histories}
\label{sec:history}
In this section, we examine the possible thermal histories of our model, with particular emphasis on how the new particle $\phi$ affects the final relic abundance of the WIMP. 

We assume a standard cosmology (radiation domination) with a reheating temperature well above $m_\phi$. We further assume that the initial number density of $\phi$ is negligible. Since $\chi$ is a WIMP, it remains in thermal equilibrium with the SM bath in the early Universe through pair annihilations. We stay agnostic about the underlying interactions between $\chi$ and the SM particles and parametrize the relevant particle physics in terms of the thermally averaged annihilation cross section, $\langle \sigma v \rangle$. Meanwhile, the  partner particle, $\phi$, which interacts only with $\chi$, is produced via the inverse decay process $\chi\chi \rightarrow \phi$ and the pair annihilation process $\chi\chi \rightarrow \phi\phi$. Depending on the Yukawa coupling $\lambda$, $\phi$ may or may not reach equilibrium with $\chi$, and hence with the bath.

The produced $\phi$ particles are unstable and eventually decay into WIMPs. If this decay occurs before the freeze-out of the WIMPs from equilibrium abundance, while the WIMPs are still in chemical equilibrium with the bath, the injected $\chi$ particles rapidly equilibrate with the bath. Consequently, chemical equilibrium erases any impact of $\phi$ on the thermal history and final abundance of $\chi$. Therefore, the thermal history in this case is simply that of the standard WIMP freeze-out from equilibrium.  

To affect the thermal history of $\chi$, $\phi$ must decay after $\chi$ has departed from equilibrium with the thermal bath; in this regime, the decay can modify the final DM abundance. Since $m_\phi>m_\chi$, if $\phi$ reaches equilibrium with $\chi$ and consequently with the bath, when $\chi$ starts to decouple from the bath, $\phi$'s number density is extremely suppressed due to decay and following its equilibrium number density. Therefore, to have any noticeable impact, $\phi$ should be produced out of thermal equilibrium and decay when $\chi$ is departing from equilibrium.
If $\phi$ decays after WIMP freeze-out from its equilibrium
abundance, its impact is determined by the number of $\chi$ particles produced and by the annihilation cross section of $\chi$ into SM particles. The $\chi$ particles produced from $\phi$ decay can annihilate among themselves or with the existing $\chi$ population, initiating a period of re-annihilation\footnote{While re-annihilation has previously been studied in a hidden sector, in this study it takes place within the same sector.}~\cite{Cheung:2010gj,Chu:2011be}, during which the WIMP yield follows a quasi-equilibrium value. Eventual freeze-out of $\chi$ from re-annihilation sets its final relic abundance. 

If re-annihilation is inefficient, the $\chi$ particles produced from $\phi$ decay simply add to the existing DM abundance left over from standard freeze-out from equilibrium, thereby increasing the final DM abundance. This can be understood as a freeze-in channel for production of WIMPs which follows their thermal freeze-out from equilibrium.

The relevant number changing processes involved in setting the WIMP  abundance are: 
\begin{itemize}
    \item ${\rm B}\,{\rm B}\leftrightarrow \chi\chi$, where B is a particle in the bath that interacts weakly with WIMPs,
    \item $\chi\chi\leftrightarrow \phi$, 
    \item $\chi\chi\leftrightarrow \phi\phi$ (t- and u-channels).
\end{itemize}
By comparing the number of reactions per unit volume per unit time in $\chi\chi\rightarrow\phi\phi$, and $\chi\chi\rightarrow\phi$, we find $\Gamma_{\chi\chi\rightarrow\phi\phi}/\Gamma_{\chi\chi\rightarrow\phi}\sim \lambda^2(T/m_\phi)^2$. Even for a small coupling, at sufficiently high temperatures, annihilation can dominate over inverse decay. However, it does not dominate the final abundance of $\phi$, since the bulk of $\phi$ production through freeze-in is IR-dominated and is not sensitive to the highest temperature of the bath~\cite{Hall:2009bx}, as long as the highest temperature is above all the relevant masses~\cite{Boddy:2024vgt}. The bulk of the production occurs when the bath temperature is of order $m_\phi$, where inverse decay is the dominant production channel. Therefore, in our analysis, we include only the inverse decay and decay processes in the evolution of the number density of $\phi$.

In our setup, the expansion of the Universe is driven by a bath of radiation with temperature $T$ consisting of $g_\star$ relativistic degrees of freedom; the Hubble expansion rate is therefore $H=\sqrt{4\pi^3/45 }\,g_\star^{1/2}T^2/m_{\rm Pl}$. We introduce the dimensionless parameter $x=m_\chi/T$ to track cosmological time. By normalizing the number density of particles, $n$, by the entropy density of the Universe, $s=(2\pi^2/45)g_{\star,S}T^3$, we define the yield of particles: $Y=n/s$. Throughout this study, we assume $g_\star=g_{\star,S}=106.75$, and we ignore the contribution of $\chi$ to $g_\star$ at early times. We further assume $dg_\star/dT=0$. Then the Boltzmann equations governing the evolution of the yield of particles are given by 
\begin{eqnarray}
 \nonumber   \frac{dY_\chi}{dx}&=&-\frac{\langle\sigma v\rangle s(x)}{xH(x)}\left(Y_\chi^2-Y_{\chi,\rm{eq}}^2\right)-2\frac{\langle\Gamma_\phi\rangle}{xH(x)}\left(Y_{\phi,{\rm eq}}\frac{Y_\chi^2}{Y_{\chi,\rm{eq}}^2}-Y_\phi\right),\\
    \frac{dY_\phi}{dx}&=&\frac{\langle\Gamma_\phi\rangle}{xH(x)}\left(Y_{\phi,{\rm eq}}\frac{Y_\chi^2}{Y_{\chi,\rm{eq}}^2}-Y_\phi\right),
    \label{eq:Boltzmanneq}
\end{eqnarray}
where $\langle \sigma v\rangle$ is the thermally-averaged annihilation cross section of the WIMPs, which in our study is parameterized as a constant assuming $s$-wave annihilation. $\langle \Gamma_\phi\rangle$ is the thermally averaged decay width of $\phi$ into $\chi$'s, given by
\begin{equation}
    \langle \Gamma_\phi\rangle=\Gamma_{\phi}\frac{K_1(m_\phi/T)}{K_2(m_\phi/T)},
    \quad
    \Gamma_{\phi} =\frac{\lambda^2}{16\pi}m_\phi\left(1-\frac{4m^2_\chi}{m^2_\phi}\right)^{3/2},
    \label{eq:decaywidth}
\end{equation}
where $\Gamma_{\phi}$ is the decay width of $\phi$ in its rest frame, and $K_i$ is the modified Bessel function of the second kind. The equilibrium number density of the  $i$th particle with mass of $m_i$ and $g_i$ degrees of freedom at temperature $T$ is given by $n_{i,{\rm eq}}=[g_i/(2\pi^2)]m_i^2TK_2(m_i/T)$. The factor of 2 in the decay term of the yield equation for $\chi$ corresponds to the production of two $\chi$ particles per decay of $\phi$.

We use the integrated Boltzmann equations, which describe the evolution of the yield rather than the full phase-space distribution. This is allowed as long as all the particles are in kinetic equilibrium with the thermal bath, i.e., $f_i(t,E)=[n_i(t)/n _{i,{\rm eq}}(t)]f_{i,{\rm eq}}(t,E)$. Although this assumption does not apply to $\phi$ when its coupling to the bath (via the WIMP) is feeble, as shown in Ref.~\cite{Du:2021jcj}, for sufficiently small decay width of $\phi$ the results of the integrated Boltzmann equations are in good agreement with those from the Boltzmann equations at the level of the phase-space distribution. In a sense, since the $\phi$ particles are produced from the thermal bath, their momenta are expected to be of the order of the bath temperature. One can therefore think of the bath temperature as an effective temperature for these particles which are produced out of equilibrium~\cite{Abdelrahim:2025fiz}.
\begin{figure}[t]
    \centering
    \includegraphics[width=\textwidth]{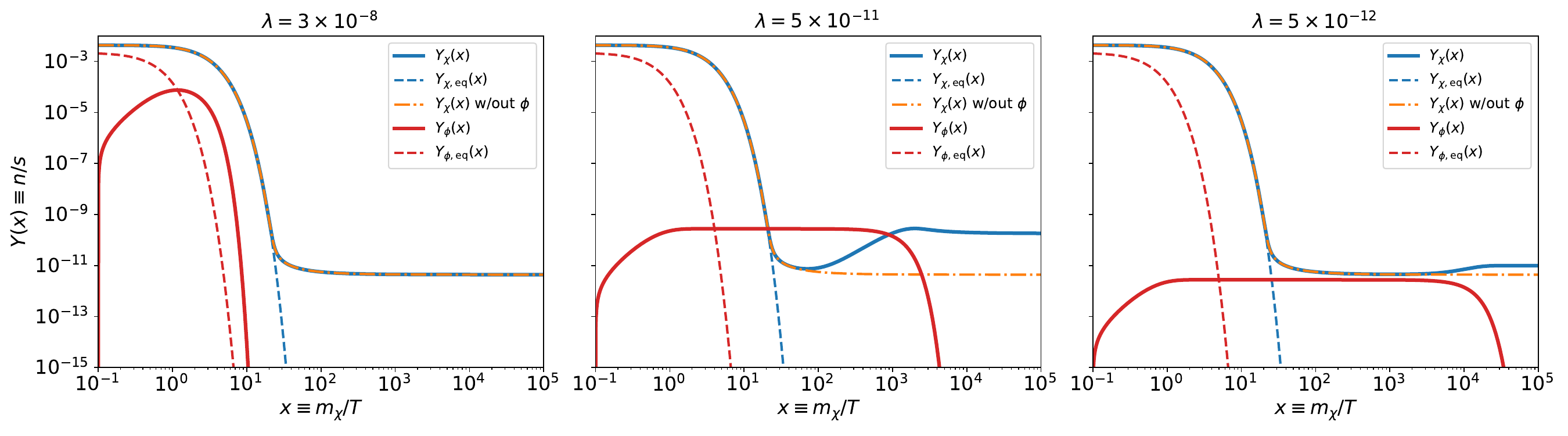}
    \caption{Evolution of the yields of a WIMP DM $\chi$ with mass $m_\chi = 100~\gev$ and annihilation cross section $\langle\sigma v\rangle_{\rm th}$, and its partner $\phi$ with mass $m_\phi = 500~\gev$, interacting through a coupling $\lambda$, as a function of $x \equiv m_\chi/T$. The solid (dashed) blue line shows the yield (equilibrium yield) of $\chi$, while the solid (dashed) red line depicts the yield (equilibrium yield) of $\phi$. The dash-dotted orange line shows the WIMP yield in the absence of a partner, whose final abundance is determined by the standard freeze-out mechanism and is matched to the observed value. Each panel is associated with a different coupling and corresponds to a distinct thermal history of the WIMPs. The left panel corresponds to $\lambda=3\times10^{-8}$ which allows the $\phi$ to decay before the WIMP freeze-out from equilibrium and the yield of DM follows the standard freeze-out. In the middle panel, $\lambda= 5\times10^{-11}$ and the decay of $\phi$ leads to a re-annihilation phase. The right panel, with $\lambda= 5\times10^{-12}$, corresponds to the case where WIMPs injected by $\phi$ freeze in and contribute to the final abundance. In the middle and right panels, DM overproduction is evident, which requires an annihilation cross section larger than $\langle\sigma v\rangle_{\rm th}$.}
    \label{fig:gamma_comparison}
\end{figure}

To illustrate how the presence of a WIMP partner can modify the thermal history of WIMPs, Fig.~\ref{fig:gamma_comparison} shows the evolution of the yields of a WIMP DM $\chi$ with mass $m_\chi = 100\gev$ and annihilation cross section $\langle\sigma v\rangle_{\rm th}$, and its partner $\phi$ with mass $m_\phi = 500\gev$, interacting through a coupling $\lambda$. The solid (dashed) blue line shows the evolution of the yield (equilibrium yield) of $\chi$, while the solid (dashed) red line shows the evolution of the yield (equilibrium yield) of $\phi$. The dash-dotted orange line shows the evolution of the WIMP yield in the absence of a partner, whose final abundance is determined by the standard freeze-out mechanism and is matched to the observed value. Each panel of Fig.~\ref{fig:gamma_comparison} corresponds to a different value of the coupling constant, illustrating three distinct thermal histories. In the left panel, $\lambda = 3\times10^{-8}$, and $\phi$ decays before the WIMP freezes out from equilibrium. Consequently, the evolution is essentially identical to the standard freeze-out scenario. In the middle panel, corresponding to $\lambda = 5\times10^{-11}$, $\phi$ decays after the WIMP freezes out from equilibrium, leading to a re-annihilation phase. Finally, in the right panel, corresponding to $\lambda = 5\times10^{-12}$, the WIMPs produced from the decay of $\phi$ do not undergo re-annihilation. Instead, they simply add to the final dark matter abundance, analogous to a freeze-in contribution. It is clear from both the middle and right panels that the late decay of the partner leads to an overproduction of WIMPs. Consequently, reproducing the observed dark matter abundance requires an annihilation cross section larger than the thermal value, $\langle\sigma v\rangle_{\rm th}$.

Motivated by these three distinct thermal histories, we now investigate each scenario in detail by solving the Boltzmann equations governing the evolution of the yields of $\chi$ and $\phi$, both analytically and numerically. We determine the final WIMP abundance, identify the range of the coupling $\lambda$ for which each thermal history is realized, and, when applicable, quantify the enhancement of the annihilation cross section relative to $\langle\sigma v\rangle_{\rm th}$ required to reproduce the observed dark matter relic abundance. 
\subsection{Standard Freeze-out from Equilibrium}
If the partner particle decays prior to WIMP freeze-out from equilibrium, i.e., if $\Gamma_\phi >H(x_{f, {\rm ann.}})$ where $x_{f,{\rm ann.}}$ denotes the time of WIMP freeze-out from equilibrium, it leaves no impact on the WIMP thermal history. In this case, the yield of WIMPs is described by the following equation:
\begin{equation}
    \frac{dY_\chi}{dx}=-\frac{\langle\sigma v\rangle s(x)}{xH(x)}\left(Y_\chi^2-Y_{\chi,\rm{eq}}^2\right),
    \quad
    \Gamma_\phi >H(x_{f,{\rm ann.}}).
    \label{eq:Yfo}
\end{equation}
For $s$-wave annihilation, this can be written as
\begin{equation}
    \frac{dY_\chi}{dx}=-\alpha x^{-2}\left(Y_\chi^2-Y_{\chi,\rm{eq}}^2\right),
    \quad
    \Gamma_\phi >H(x_{f,{\rm ann.}}),
    \label{eq:Yfo_alpha}
\end{equation}
with
\begin{equation}
    \alpha\equiv \sqrt{\frac{\pi}{45}}\frac{g_\star}{\sqrt{g_{\star,S}}}m_{\rm Pl}m_\chi\langle\sigma v\rangle.
    \label{eq:alpha}
\end{equation}

Adopting the semi-analytical treatment of Ref.~\cite{Kolb:1990vq}, the freeze-out time of the WIMP is estimated as
\begin{equation}
   x_{f,{\rm ann.}}\simeq\ln\,[(2+c_1)\alpha a c_1]-\frac{1}{2}\ln\,\{\ln\,[(2+c_1)\alpha a c_1]\},
   \label{eq:xfann}
\end{equation}
where $c_1$ is a numerical constant of order unity, and the constant $a$ is defined as
\begin{equation}
a\equiv \frac{45}{4\sqrt{2}\pi^{7/2}}\frac{g_\chi}{g_\star}.
\end{equation}
The final yield of DM, set by freeze-out from equilibrium, is given by
\begin{equation}
    Y_\infty=\frac{x_{f,{\rm ann.}}}{\alpha}=\sqrt{\frac{45}{\pi}}\frac{\sqrt{g_{\star,S}}}{g_\star}\frac{x_{f,{\rm ann.}}}{m_{\rm Pl}m_\chi\langle\sigma v\rangle}.
    \label{eq:yieldfroutEQ}
\end{equation}
Requiring that $\phi$ decays before the WIMPs freeze out from equilibrium determines the minimum coupling between the partner particle and the WIMP, above which the standard WIMP thermal history remains unchanged. This coupling is
determined by the condition $x_{\rm dec.}\lesssim x_{f,{\rm ann.}}$ where $x_{\rm dec.}$ denotes the decay time of $\phi$ obtained from $H(x_{\rm dec.})\sim \Gamma_{\phi}$. For couplings below this threshold, a re-annihilation phase may emerge. We denote this coupling by $\lambda^{\rm max}_{\rm re\text{-}ann.}$, which can be estimated as
\begin{equation}
\lambda^{\rm max}_{\rm re\text{-}ann.}\simeq\frac{4\sqrt{2}}{\sqrt{3}} \left(\frac{\pi^5}{5}\right)^{1/4} g^{1/4}_\star x_{f,{\rm ann.}}^{-1}\frac{m_\chi}{\sqrt{m_{\rm Pl}m_\phi}}\left(1-\frac{4m^2_\chi}{m^2_\phi}\right)^{-3/4}.
\end{equation} 
We note that this estimate assumes the instantaneous decay of the $\phi$ particles at $t=\Gamma_\phi^{-1}$. A more accurate estimate can be obtained by requiring that most of the $\phi$ particles decay before freeze-out from equilibrium. For example, demanding that $95\%$ of the $\phi$ particles have decayed corresponds to requiring that freeze-out from equilibrium occurs at $t\gtrsim 3\Gamma_\phi^{-1}$.
\subsection{Freeze-out from Re-annihilation}
For $\lambda < \lambda^{\rm max}_{\rm re\text{-}ann.}$, the non-thermal $\chi$ particles produced from the decay of $\phi$ will not thermalize with the bath. However, they increase the comoving number density of WIMPs and, depending on the lifetime of the partner particle, the WIMP abundance may become sufficiently large for annihilations into bath particles to become efficient again, thereby initiating a re-annihilation phase. Since, at this stage, the temperature of the bath is much smaller than the WIMP mass, the WIMP abundance cannot follow its equilibrium abundance and instead tracks a ``quasi-equilibrium" value. At this time, WIMP production from the bath and partner production via inverse decay are suppressed, and $Y_{\chi,\rm{eq}}\ll Y_\chi$ and $Y_{\phi,\rm{eq}}\ll Y_{\phi}$. Then the Boltzmann equation for $\chi$ can be approximated as
\begin{eqnarray}
    \nonumber \frac{dY_\chi}{dx}&\simeq&-\frac{\langle\sigma v\rangle s(x)}{xH(x)}Y_\chi^2+2\frac{\langle\Gamma_\phi\rangle}{xH(x)}Y_\phi\\
\nonumber     &=&-\frac{\langle\sigma v\rangle s(x)}{xH(x)}\left(Y^2_\chi-Y^2_{\chi,{\rm QE}}\right)\\
     &=&-\alpha x^{-2}\left(Y^2_\chi-Y^2_{\chi,{\rm QE}}\right),
     \label{eq:reanneq}
\end{eqnarray}
 where the quasi-equilibrium value is defined by
\begin{equation}
    Y_{\chi,{\rm QE}}\equiv\left[2\frac{\langle\Gamma_\phi\rangle}{\langle\sigma v\rangle s(x)}Y_\phi\right]^{1/2},
    \label{eq:YQE}
\end{equation}
with $\alpha$ given by Eq.~(\ref{eq:alpha}).

The re-annihilation eventually ends, and DM freezes out to its final abundance. Since Eq.~(\ref{eq:reanneq}) has a form similar to Eq.~(\ref{eq:Yfo_alpha}), we apply a semi-analytical approach similar to the one used in Ref.~\cite{Kolb:1990vq} to estimate the time of freeze-out from quasi-equilibrium and the final DM yield. Before applying this approach, we first need to estimate the yield of $\phi$ as a function of time, since it determines $Y_{\chi,{\rm QE}}$. Since $\phi$ is produced through freeze-in, its abundance can be estimated as follows: At early times, $\phi$ production is dominated by inverse decays from the bath, and the decay term can therefore be neglected. Assuming $Y_\chi=Y_{\chi,{\rm eq}}$, we can solve the Boltzmann equation governing the evolution of $Y_\phi$ to obtain the abundance of $\phi$ produced through inverse decays, $Y^{\rm inv.}_\phi(x)$, 
\begin{equation}
    Y^{\rm inv.}_\phi(x)=\int^x_0 \frac{\langle\Gamma_\phi\rangle}{x'H(x')}Y_{\phi,{\rm eq}}(x')dx'= \frac{135\sqrt{5}}{8\pi^{11/2}}\frac{g_\phi}{\sqrt{g_\star}g_{\star S}}\frac{m^2_\phi m_{\rm pl}\Gamma_{\phi}}{m_\chi^4}\int^x_0x'^3 K_1\left(\frac{m_\phi}{m_\chi}x'\right) dx',\quad x\lesssim x_{\rm dec.}.
    \label{eq:Yphi1}
\end{equation}
The yield of $\phi$ saturates at its maximum value around $x \sim m_\chi/m_\phi$. This saturated value is insensitive to the subsequent evolution at larger values of $x$ and can therefore be approximated as $Y^{\rm inv.}_\phi(\infty)$.
At later times, the $\phi$ abundance is depleted due to its decay and can therefore be estimated as
\begin{equation}
  Y_\phi(x)= Y^{\rm inv.}_\phi(\infty)\exp\left[-\Gamma_{\phi}/\left(2H(x)\right)\right]= \frac{405\sqrt{5}}{16\pi^{9/2}}\frac{g_\phi}{\sqrt{g_\star}g_{\star S}}\frac{m_{\rm Pl}\Gamma_{\phi}}{m_\phi^2} \exp\left[-\Gamma_{\phi}/\left(2H(x)\right)\right],
  \quad x\gtrsim 6m_\chi/m_\phi.
  \label{eq:Yphi2}
\end{equation}
By using a semi-analytical approach which has been outlined in Appendix~\ref{appx:reanni}, we can show that DM freezes out from re-annihilation at $x_{f,{\rm re\text{-}ann.}}$, given by
\begin{eqnarray}
  \nonumber  x_{f,{\rm re\text{-}ann.}}&\simeq& \frac{\sqrt{2}\pi^{3/4}}{5^{1/4}\sqrt{3}}g_\star^{1/4}\frac{m_\chi}{\sqrt{m_{\rm Pl}\Gamma_{\phi}}}\sqrt{-\ln z}\left[1-\frac{\ln(-\ln z)}{2\ln z}\right]\\
    &=& x_{\rm dec.}\sqrt{-\ln z}\left[1-\frac{\ln(-\ln z)}{2\ln z}\right],
\end{eqnarray}
with
\begin{equation}
    z\equiv \frac{8\pi^{19/2}b^4}{675\sqrt{5}c_2^4(2+c_2)^4}\frac{g^6_{\star,S}}{g_\phi^2g^{7/2}_\star}\frac{m_\phi^4}{m^5_{\rm Pl}\langle\sigma v\rangle^2\Gamma^3_{\phi}},
\end{equation}
where $b$ and $c_2$ are numerical factors of order unity. Due to the logarithmic dependence of $x_{f,{\rm re\text{-}ann.}}$ on these parameters, their precise values do not play an important role. It is easy to show that $x_{f,{\rm re\text{-}ann.}}= {\rm few}\times x_{\rm dec.}$.

\begin{figure}[t]
   \centering
    \includegraphics[width=0.5\textwidth]{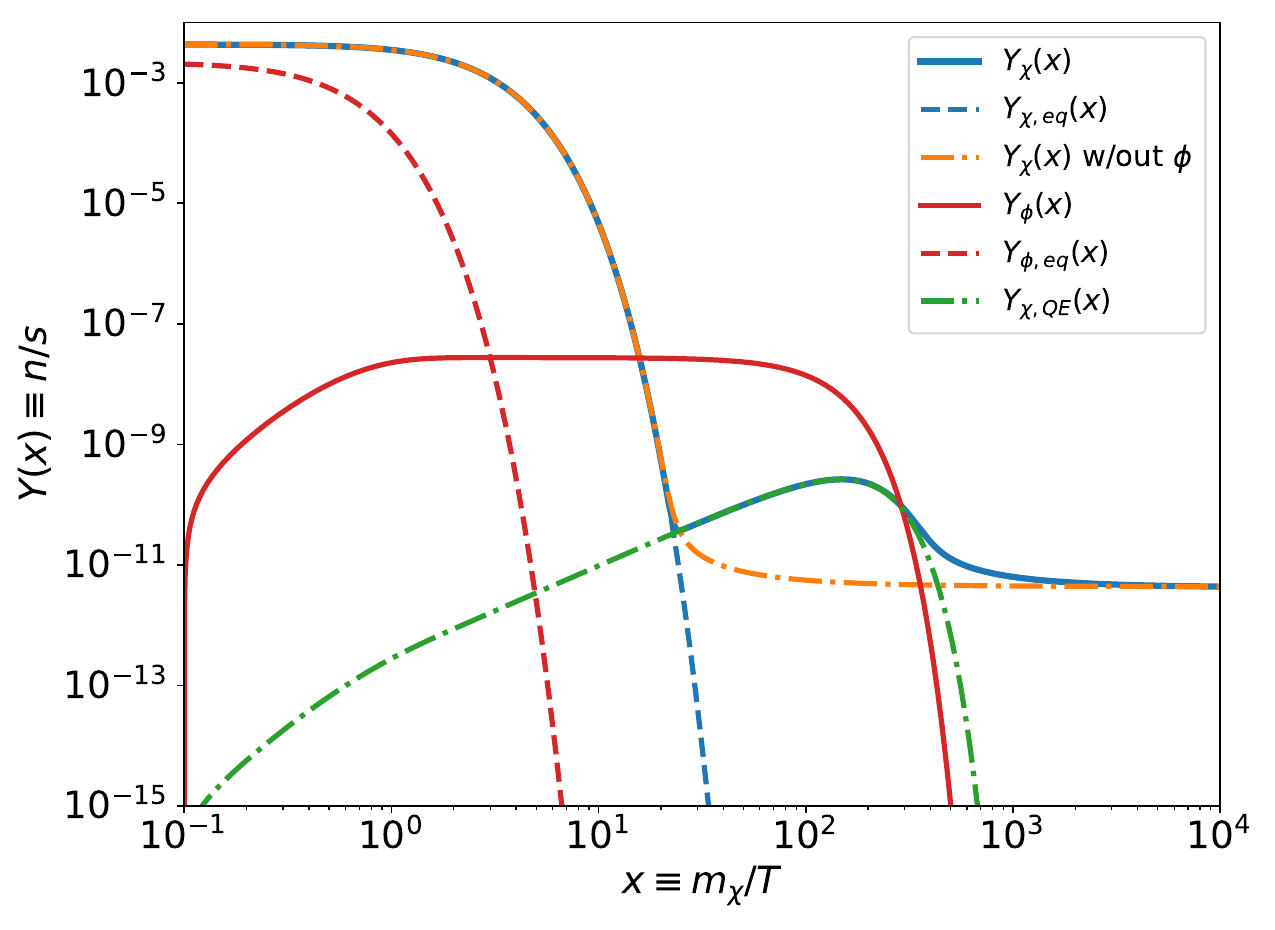}
   \caption{Evolution of the yields of a WIMP DM and its partner for $\{m_\chi = 100~\gev, m_\phi = 500~\gev,\lambda=5\times 10^{-10}\}$. The solid (dashed) blue lines show $Y_\chi(x)$ ($Y_{\chi,{\rm eq}}(x)$) and the solid (dashed) red lines shows $Y_\phi(x)$ ($Y_{\phi,{\rm eq}}(x)$). The annihilation cross section is set to $\langle\sigma v\rangle= 2.78\times 10^{-25}~{\rm cm^3\,s^{-1}}\sim 10\langle\sigma v\rangle_{\rm th}$ to match the observed relic abundance of DM today. The dash-dotted orange line depicts the evolution of $Y_\chi(x)$ when in the absence of a partner when annihilation cross section is $\langle\sigma v\rangle_{\rm th}$. The quasi-equilibrium yield, $Y_{\chi, \rm QE}(x)$ is displayed as the dash-dotted green line. The DM yield follows the quasi-equilibrium value during the re-annihilation phase and eventually freezes out to set its final abundance.}
    \label{fig:yield_plot}
\end{figure}

Not surprisingly, the final DM yield after freeze-out from re-annihilation is given by the same formula as the yield after freeze-out from equilibrium, Eq.~(\ref{eq:yieldfroutEQ}), with $x_{f,{\rm ann.}}$ replaced by $x_{f,{\rm re\text{-}ann.}}$:
\begin{equation}
   Y_\infty=\frac{x_{f,{\rm re\text{-}ann.}}}{\alpha}.
\end{equation}
We note that the final DM abundance is proportional to $x_{f,{\rm re\text{-}ann.}}$. Since, in the re-annihilation scenario, freeze-out occurs later than in the standard freeze-out from equilibrium, i.e., $x_{f,{\rm ann.}}<x_{\rm dec.}<x_{f,{\rm re\text{-}ann.}}$, the final abundance of WIMPs after re-annihilation is expected to be larger than that obtained from standard freeze-out. Therefore, in the presence of a partner particle that induces re-annihilation, reproducing the observed DM abundance requires an annihilation cross section larger than the canonical value. The enhancement in annihilation cross section is given by
\begin{equation}
    \frac{\langle\sigma v\rangle_{\rm re\text{-}ann.}}{\langle\sigma v\rangle_{\rm ann.}}=\frac{x_{f,{\rm re\text{-}ann.}}}{x_{f,{\rm ann.}}}\simeq{\rm few}\times\frac{x_{\rm dec.}}{x_{f,{\rm ann.}}}\sim\frac{1}{\lambda}\frac{m_\chi}{\sqrt{m_{\rm Pl}m_{\phi}}}.
    \label{eq:enhanSV}
\end{equation}
We also point out that $x_{f,{\rm re\text{-}ann.}} \propto \Gamma_{\phi}^{-1/2}$, implying that, in the re-annihilation regime, partner particles with longer lifetimes require larger WIMP annihilation cross sections in order to reproduce the observed DM abundance today. Finally, it is worth mentioning that while, in the standard freeze-out scenario from equilibrium, the freeze-out time depends only mildly (logarithmically) on the DM mass, c.f. Eq.~(\ref{eq:xfann}), in the re-annihilation case the freeze-out time scales linearly with the DM mass, up to the dependence of the partner particle decay width on the WIMP mass. In both cases, however, the freeze-out time depends only logarithmically on the annihilation cross section. Fig.~\ref{fig:yield_plot} shows the evolution of the DM and partner particle yields obtained by numerically solving the Boltzmann equations for the choice of parameters $\{m_\chi=100~{\rm GeV},m_\phi=500~{\rm GeV}, \lambda=5\times10^{-10}\}$, which gives rise to a re-annihilation phase. The annihilation cross section required to explain the observed abundance of DM today is calculated to be $\langle\sigma v\rangle = 2.78\times 10^{-25}~{\rm cm^3\,s^{-1}}$, which is $\sim$ten times larger than $\langle\sigma v\rangle_{\rm th}$. The quasi-equilibrium yield is shown as the dash-dotted green line. As expected, the DM yield (solid blue line) tracks the quasi-equilibrium value during the re-annihilation phase and eventually freezes out, setting its final abundance. For reference, the dash-dotted orange line is the evolution of the WIMP yield when it has no partner and its annihilation cross section is $\langle\sigma v\rangle_{\rm th}$.

By reducing the coupling $\lambda$ to smaller values, re-annihilation eventually becomes less efficient. Although some residual annihilation may still occur, the DM yield no longer tracks the quasi-equilibrium value, and therefore there is no subsequent freeze-out from this state. Thus, there exists a minimum coupling, denoted by $\lambda^{\rm min}_{\rm re\text{-}ann.}$, below which the re-annihilation phase no longer takes place. In this regime, the WIMPs injected through the decay of $\phi$ simply add to the pre-existing population of WIMPs from freeze-out from equilibrium, contributing to the final relic abundance as a freeze-in component. The threshold coupling, $\lambda^{\rm min}_{\rm re\text{-}ann.}$, can be estimated by imposing the condition $z(\lambda^{\rm min}_{\rm re\text{-}ann.})=1/e$, which is equivalent to requiring that the DM yield never tracks the quasi-equilibrium value and is therefore not set by freeze-out from it. In other words, for $z(\lambda)>1/e$, there is no physical value of $x_{f,\rm re\text{-}ann.}$ at which the DM yield becomes comparable to the quasi-equilibrium yield; see Appendix~\ref{appx:reanni} for details. By further requiring the DM yield to match the observed value, $Y_{\rm CDM}\simeq 4.4\times10^{-10}(1~\rm GeV/m_\chi)$, one can solve for the threshold coupling, 
\begin{eqnarray}
   \nonumber \lambda^{\rm min}_{\rm re\text{-}ann.}&\sim& \frac{4\sqrt{2}\pi^{11/4}e^{1/4}}{3\times 5^{3/4}}\frac{g_{\star,S}^{5/4}}{\sqrt{g_\phi}\sqrt{g_{\star}}}\sqrt{Y_{\rm CDM}}\sqrt{\frac{m_\phi}{m_{\rm Pl}}}\left(1-\frac{4m^2_\chi}{m^2_\phi}\right)^{-3/4}\\
    &\simeq& 3.5\times 10^{-4}\frac{g_{\star,S}^{5/4}}{\sqrt{g_\phi}\sqrt{g_{\star}}}\sqrt{\frac{m_\phi}{m_{\rm Pl}}}\sqrt{\frac{1~\rm GeV}{m_\chi}}\left(1-\frac{4m^2_\chi}{m^2_\phi}\right)^{-3/4}. 
    \label{eq:lambda_threshold}
\end{eqnarray}
For a weak-scale partner, one finds $\lambda^{\rm min}_{\rm re\text{-}ann.}\sim \mathcal{O}(10^{-12}-10^{-11})$.

The maximum enhancement in annihilation cross section corresponds to $\lambda^{\rm min}_{\rm re\text{-}ann.}$, and is given by
\begin{eqnarray}
  \nonumber  \left(\frac{\langle\sigma v\rangle_{\rm re\text{-}ann.}}{\langle\sigma v\rangle_{\rm ann.}}\right)_{\rm max}&=&\frac{\langle\sigma v\rangle_{\rm re\text{-}ann.}(\lambda^{\rm min}_{\rm re\text{-}ann.})}{\langle\sigma v\rangle_{\rm ann.}}=\frac{x_{f,{\rm re\text{-}ann.}}(\lambda^{\rm min}_{\rm re\text{-}ann.})}{x_{f,{\rm ann.}}}\simeq \frac{x_{\rm dec.}(\lambda^{\rm min}_{\rm re\text{-}ann.})}{x_{f,{\rm ann.}}}\\
   \nonumber &\simeq&
    \frac{\sqrt{15}}{20\times e^{1/4}\pi^{3/2}}\frac{\sqrt{g_\phi}g_\star^{3/4}}{g_{\star,S}^{5/4}}\frac{1}{\sqrt{Y_{\rm CDM}}}\frac{m_\chi}{m_\phi}\\
    &\simeq&1.3\times 10^3 \frac{\sqrt{g_\phi}g_\star^{3/4}}{g_{\star,S}^{5/4}}\frac{m_\chi}{m_\phi}\sqrt{\frac{m_\chi}{1~\rm GeV}},
\end{eqnarray}
where we take $x_{f,{\rm ann.}}\simeq20$. For a weak-scale partner not much heavier than the WIMP, we therefore expect $\left(\langle\sigma v\rangle_{\rm re\text{-}ann.}/\langle\sigma v\rangle_{\rm ann.}\right)_{\rm max}\sim \mathcal{O}(10^2-10^4)$.
\subsection{Freeze-out Followed by Freeze-in}
For $\lambda <\lambda^{\rm min}_{\rm re\text{-}ann.}$, the WIMP initially freezes out from equilibrium at $x_{f,{\rm ann.}}$, with a yield $Y_\chi(x_{f,{\rm ann.}})=x_{f,{\rm ann.}}/\alpha$. Later, $\phi$ decays and injects WIMPs into the bath, adding to the existing $Y_\chi(x_{f,{\rm ann.}})$ without significant re-annihilation, i.e., not enough to cause the yield to track the quasi-equilibrium value. This process can be understood as a freeze-in production channel for WIMPs. Thus, the freeze-out from equilibrium followed by freeze-in from the decay of the partner particle determines the final WIMP abundance. In this case, the Boltzmann equation governing the evolution of $\chi$ reduces to
\begin{equation}
    \frac{dY_\chi}{dx} \simeq 2\frac{\langle \Gamma_\phi \rangle}{x H(x)} Y_\phi,
\quad 
Y_\chi(x_{f,{\rm ann.}})=\frac{x_{f,{\rm ann.}}}{\alpha},
\quad
x>x_{f,{\rm ann.}},
\end{equation}
where $Y_\phi$ is given by Eq.~(\ref{eq:Yphi2}), and $x_{f,{\rm ann.}}/\alpha$ fixes the initial condition of the differential equation. The final DM yield is then given by
\begin{equation}
    Y_\infty=Y_\chi(x_{f,{\rm ann.}})+Y^{\rm f.i.}_\chi,
\end{equation}
with
\begin{equation}
    Y^{\rm f.i.}_{\chi}=\frac{405\sqrt{5}}{8\pi^{9/2}}\frac{g_\phi}{g_{\star,S}\sqrt{g_\star}}\frac{m_{\rm Pl}\Gamma_\phi}{m^2_\phi}\exp\left(-\frac{3\sqrt{5}}{4\pi^{3/2}}\frac{x^2_{f,{\rm ann.}}}{\sqrt{g_\star}}\frac{m_{\rm Pl}\Gamma_\phi}{m^2_\chi}\right)= 2 Y^{\rm inv.}_\phi(\infty),
\end{equation}
where the exponent is a very small number such that the exponential is approximately one. This equation simply reflects the fact that the yield of $\chi$ particles produced from $\phi$ decay is twice the yield of $\phi$ particles produced via inverse decay.

Due to the additional yield contribution from freeze-in, the annihilation cross section responsible for setting $Y_\chi(x_{f,{\rm ann.}})$ must be larger than the canonical value. The enhancement in annihilation cross section due to freeze-in is given by
\begin{equation}
    \frac{\langle\sigma v\rangle_{\rm f.i.}}{\langle\sigma v\rangle_{\rm ann.}}=\frac{1}{1-\left[\alpha(\langle \sigma v\rangle_{\rm ann.})Y^{\rm f.i.}_{\chi}/x_{f,{\rm ann.}}\right]}=\frac{1}{1-\frac{135}{128\pi^5}\frac{g_\phi\sqrt{g_{\star}}}{g_{\star,S}^{3/2}}\frac{1}{x_{f,{\rm ann.}}}\frac{m_\chi m^2_{\rm Pl}\langle \sigma v\rangle_{\rm ann.}}{m_\phi}\left(1-\frac{4m^2_\chi}{m^2_\phi}\right)^{3/2}\lambda^2},
    \label{eq:sigmavfi}
\end{equation}
which reaches a maximum around
\begin{eqnarray}
 \nonumber  \lambda^{\rm max}_{\rm f.i.} &=&\frac{8\sqrt{2}\pi^{5/2}}{3\sqrt{15}} \frac{g^{3/4}_{\star,S}}{\sqrt{g_\phi}g^{1/4}_\star}\sqrt{x_{f,{\rm ann.}}}\frac{\sqrt{m_\phi}}{\sqrt{m_\chi}m_{\rm Pl}\sqrt{\langle \sigma v\rangle_{\rm ann.}}}\left(1-\frac{4m^2_\chi}{m^2_\phi}\right)^{-3/4}\\
   &=&\frac{8\sqrt{2}\pi^{11/4}}{9\times 5^{3/4}} \frac{g^{1/4}_\star \sqrt {g_{\star,S}}}{\sqrt{g_\phi}}\sqrt{Y_{\rm CDM}}\frac{\sqrt{m_\phi}}{\sqrt{m_{\rm Pl}}}\left(1-\frac{4m^2_\chi}{m^2_\phi}\right)^{-3/4},
   \label{eq:lambda_max}
\end{eqnarray}
and drops quickly for smaller couplings. In the last step we used $Y_{\rm CDM}=x_{f,{\rm ann.}}/\alpha$. The enhancement in annihilation cross section in this scenario is only noticeable if the coupling happens to be close to $\lambda^{\rm max}_{\rm f.i.}$. To better understand Eq.~(\ref{eq:lambda_max}), it is useful to note that at $\lambda=\lambda^{\rm max}_{\rm f.i.}$, the enhancement formally diverges, and the DM abundance today would be entirely explained by the freeze-in contribution from the decay of $\phi$. For the contribution from freeze-out to vanish, the annihilation cross section would have to be arbitrarily large. Of course, this is not physically possible. Instead, for a sufficiently large annihilation cross section, the injected WIMPs undergo some re-annihilation. However, the re-annihilation is not strong enough for the WIMP yield to track the quasi-equilibrium value. Therefore, the enhancement does not actually diverge at $\lambda^{\rm max}_{\rm f.i.}$. The divergent behavior of Eq.~(\ref{eq:lambda_max}) arises because, in deriving this equation, we have neglected re-annihilation for simplicity. More precisely, for $\lambda^{\rm max}_{\rm f.i.}\lesssim\lambda\lesssim \lambda^{\rm min}_{\rm re\text{-}ann.}$, the final abundance of WIMPs is determined by a combination of freeze-in and partial re-annihilation, with the re-annihilation not being efficient enough for the WIMP yield to track the quasi-equilibrium value and eventually freeze out from it. From Eqs.~(\ref{eq:lambda_threshold}) and (\ref{eq:lambda_max}) one can find that $\lambda^{\rm max}_{\rm f.i.}/\lambda^{\rm min}_{\rm re\text{-}ann.}\simeq 0.5$. Therefore, it is safe to assume that for $\lambda <\lambda^{\rm min}_{\rm re\text{-}ann.}$ the DM abundance is set by freeze-out from equilibrium followed by freeze-in. 

\begin{figure}[t]
    \centering
    \includegraphics[width=0.5\textwidth]{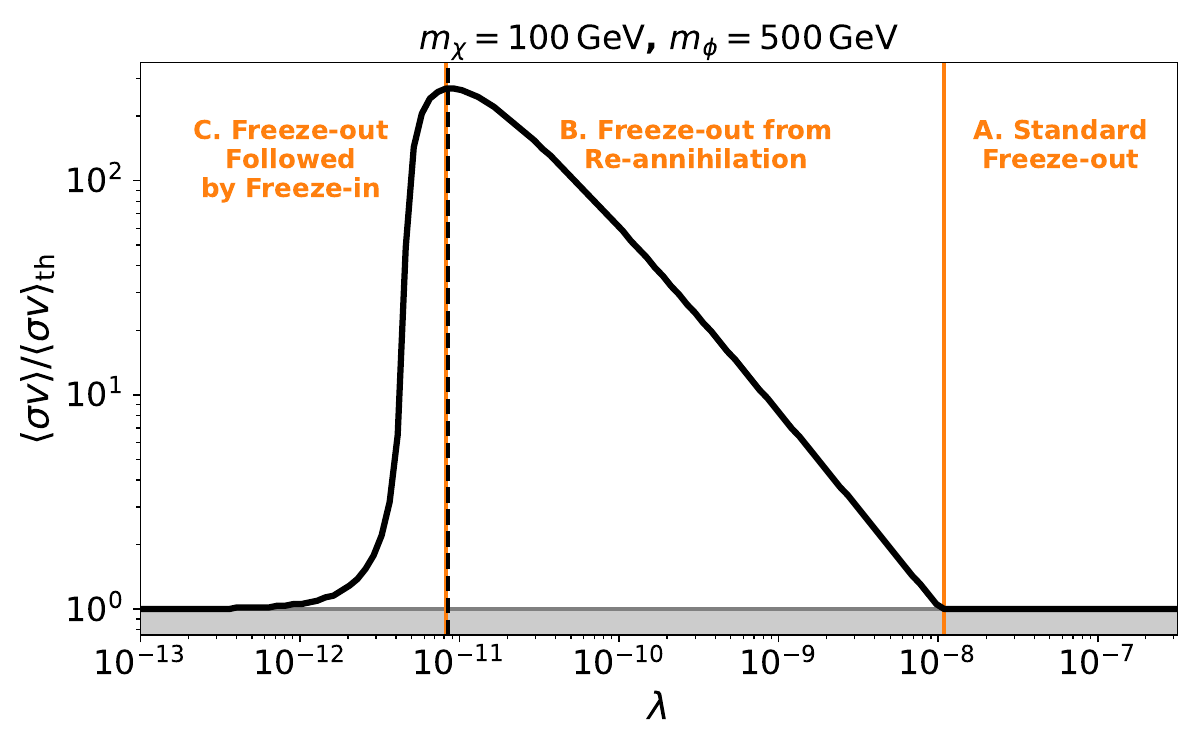}
\caption{The required WIMP annihilation cross section to explain the observed DM relic abundance today, obtained by numerically solving the Boltzmann equations and normalized to $\langle\sigma v\rangle_{\rm th}$, as a function of the coupling $\lambda$. The mass parameters are fixed to ${m_\chi=100~{\rm GeV},m_\phi=500~{\rm GeV}}$. The vertical orange lines, determined numerically, indicate the boundaries between the different thermal histories, labeled as in the Sec.~\ref{sec:history} subsections. The dashed black vertical line shows our estimate of the position of the peak, $\lambda^{\rm min}_{\rm re\text{-}ann.}.$}
    \label{fig:lambda_scan}
\end{figure}

Finally, we conclude this subsection by examining how the annihilation cross section of a WIMP DM candidate must be modified relative to the canonical value, $\langle\sigma v\rangle_{\rm th}$, to reproduce the observed DM relic abundance when the coupling between the WIMP and its partner is varied. Fig.~\ref{fig:lambda_scan} shows the required WIMP annihilation cross section calculated by numerically solving Boltzmann equations and normalized to $\langle\sigma v\rangle_{\rm th}$, as a function of the coupling $\lambda$. The mass parameters are fixed to $\{m_\chi=100~{\rm GeV},m_\phi=500~{\rm GeV}\}$. Vertical orange lines, which are calculated numerically, show the boundaries between the different thermal histories, as labeled on the $(\lambda,\langle\sigma v\rangle)$ plane. For large values of $\lambda$, i.e., $\lambda\gtrsim 10^{-8}$, the decay of the partner does not affect the thermal freeze-out of the WIMPs and therefore $\langle\sigma v\rangle/\langle\sigma v\rangle_{\rm th}=1$. By decreasing the value of the coupling, decay of the partner leads to re-annihilation and therefore requires $\langle\sigma v\rangle/\langle\sigma v\rangle_{\rm th}>1$. For these masses, the enhancement can be as large as two orders of magnitude, reaching its maximum for $\lambda\simeq 8.6 \times 10^{-12}$, as estimated by $\lambda^{\rm min}_{\rm re\text{-}ann.}$ given by Eq.~(\ref{eq:lambda_threshold}). The dashed black vertical line in  Fig.~\ref{fig:lambda_scan} marks our estimate which is in excellent agreement with numerical calculation of the peak position. By decreasing the coupling below this peak, the re-annihilation is not efficient and the injected WIMPs only add to the final abundance of the DM as a freeze-in contribution. In this region, by decreasing the coupling further, the enhancement drops quickly.  For very small couplings, we again find $\langle\sigma v\rangle/\langle\sigma v\rangle_{\rm th}=1$. It is worth noting that for a fixed set of mass parameters, two different thermal histories (freeze-out from re-annihilation and freeze-out followed by freeze-in) with different values of $\lambda$ can lead to the same enhancement in the annihilation cross section required to explain the DM abundance today. 

In our analysis, we neglect the time dependence of $g_{\star,S}$. It has been shown that for light WIMPs whose freeze-out occurs around $T\sim 100~{\rm MeV}$, where $g_{\star,S}$ changes rapidly, neglecting its variation can lead to an underestimate of the required annihilation cross section by a factor of $2$–$3$~\cite{Steigman:2012nb}. In our model, freeze-out from re-annihilation can occur at much later times, even for heavy WIMPs. We therefore expect that neglecting the time dependence of $g_{\star,S}$ also leads to an underestimate of the required annihilation cross section in this case, although the effect is expected to be of a similar order. The important distinction is that the final yield scales as $Y_\infty\propto g_\star^{-1/4}$ for freeze-out from re-annihilation, compared with $Y_\infty\propto g_\star^{-1/2}$ for the standard freeze-out from equilibrium, assuming $g_\star=g_{\star,S}$.
\subsection{Beyond the Minimal Setup: WIMP with Multiple Partners}
So far, we have investigated the impact of the decay of a single partner of the WIMP on its thermal history and final abundance. This idea can be generalized to more complicated scenarios in which the WIMP has multiple partners (that is, scalar particles that interact exclusively with the WIMPs and are sufficiently heavy to eventually decay into them). In this case, the combined effect of multiple partners can lead to a non-trivial evolution of the WIMP yield. Nevertheless, the final abundance is set by freeze-out from the quasi-equilibrium value associated with the longest-lived partner. Other possibilities, such as double re-annihilation or re-annihilation followed by freeze-in, can also arise. A detailed study of these scenarios is beyond the scope of this work, and we leave it for future work.

\section{Constraints}
\label{sec:constraints}
As we showed in the previous section, when the final abundance of the WIMP is set by freeze-out from re-annihilation or by freeze-out followed by freeze-in, a larger WIMP annihilation cross section than the canonical thermal value is required to achieve the observed DM abundance. The enhancement in the annihilation cross section, given by Eqs.~(\ref{eq:enhanSV}) and (\ref{eq:sigmavfi}), cannot be arbitrarily large and must be consistent with theoretical and observational constraints. These constraints include the perturbative unitarity bound and, depending on the annihilation final states, constraints from the cosmic microwave background (CMB), gamma rays, cosmic rays, and high-energy neutrinos. In this section, we briefly review these constraints and show that our scenario provides a well-motivated target for indirect detection searches for DM with current and upcoming gamma-ray and neutrino telescopes. We distinguish between two different final states of WIMP annihilation, visible final states (that is, any SM final state other than neutrinos, such as photons, charged leptons, and hadrons) and invisible final states (neutrinos). Our results are presented in Fig.~\ref{fig:indirect_detection}.

In general, the annihilation cross section is bounded from above by $S$-matrix unitarity, which gives, for $s$-wave annihilation of a pair of WIMPs,
$\langle \sigma v\rangle \leq 4\pi/(m_\chi^2\langle v^{-1}\rangle)$~\cite{griest:unitarity},
where $\langle v^{-1}\rangle \simeq \sqrt{x_{f,{\rm ann.}}/\pi}$~\cite{Blum:2014dca} and $x_{f,{\rm ann.}}\simeq 20$. The region excluded by the unitarity bound, which is relevant at the upper end of the WIMP mass range, is shaded in purple in Fig.~\ref{fig:indirect_detection}. The annihilation cross section of the WIMPs is also bounded from below by $\langle\sigma v\rangle_{\rm th}$, since for $\langle\sigma v\rangle<\langle\sigma v\rangle_{\rm th}$, WIMPs are overproduced and would overclose the Universe. This region is shaded in gray. 

For annihilation into visible final states, we consider the upper bounds on the annihilation cross section obtained by combining data from Planck measurements of the CMB~\cite{Planck:2015fie}, {\it Fermi}-LAT observations of dwarf spheroidal galaxies (dSphs)~\cite{Fermi-LAT:2016uux}, and AMS-02 measurements of the positron fraction~\cite{AMS:2014xys,AMS:2014bun}. This largely model-independent approach was first developed in Ref.~\cite{Leane:2018kjk}, and we adopt their result in our analysis. Updated analyses in each of these aspects have since appeared (CMB~\cite{Planck:2018vyg,Myers:2025pfx}, AMS-02 positron fraction~\cite{PhysRevLett.122.041102,AMS:2023hom}, {\it Fermi}-LAT dSphs~\cite{Circiello:2026inp}). The region excluded by constraints on annihilation to visible states is shaded in blue and labeled ``Visibles'' in the left panel of Fig.~\ref{fig:indirect_detection}. Constraints on annihilation to visible final states that are relevant at the lower end of the WIMP mass range place a lower bound of $\sim 20~\gev$ on the WIMP mass~\cite{Leane:2018kjk}. 

Within the remaining parts of the parameter space, an annihilation cross section larger than $\langle\sigma v\rangle_{\rm th}$ may still be allowed, and the main challenge in this case is to reconcile the relic abundance of the WIMP with the observed value. Our scenario can explain the observed DM abundance for different choices of the partner particle's lifetime. Current gamma-ray and neutrino telescopes are already probing parts of the allowed parameter space, while upcoming experiments are expected to extend this reach. These searches can therefore indirectly probe the presence of the WIMP's partner through its impact on the WIMP annihilation cross section. We briefly review these searches in the following subsection.
\subsection{Indirect Detection Searches}
\label{sec:indir_probes}
\begin{figure}[t]
    \centering
    \includegraphics[width=0.45\textwidth]{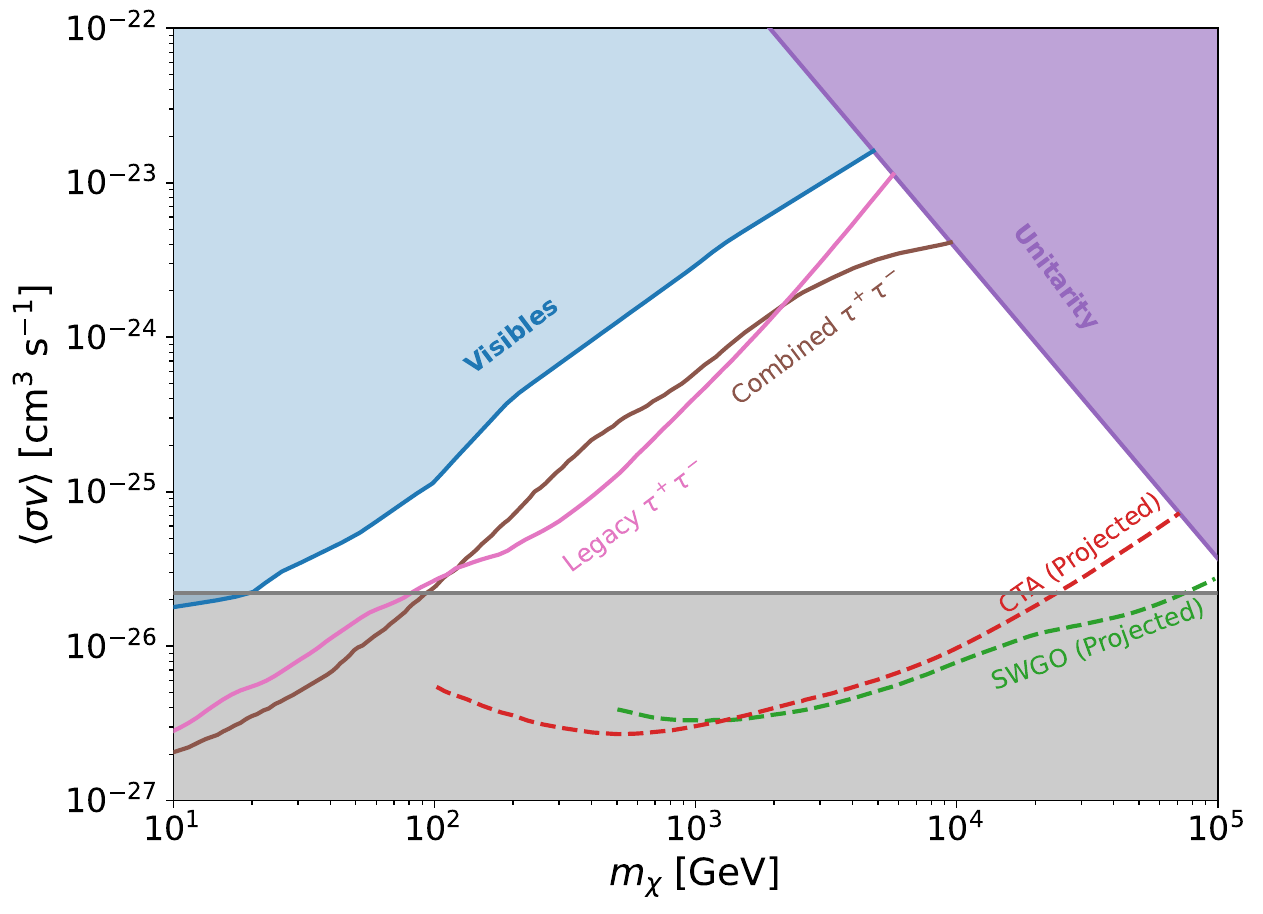}
    \includegraphics[width=0.45\textwidth]{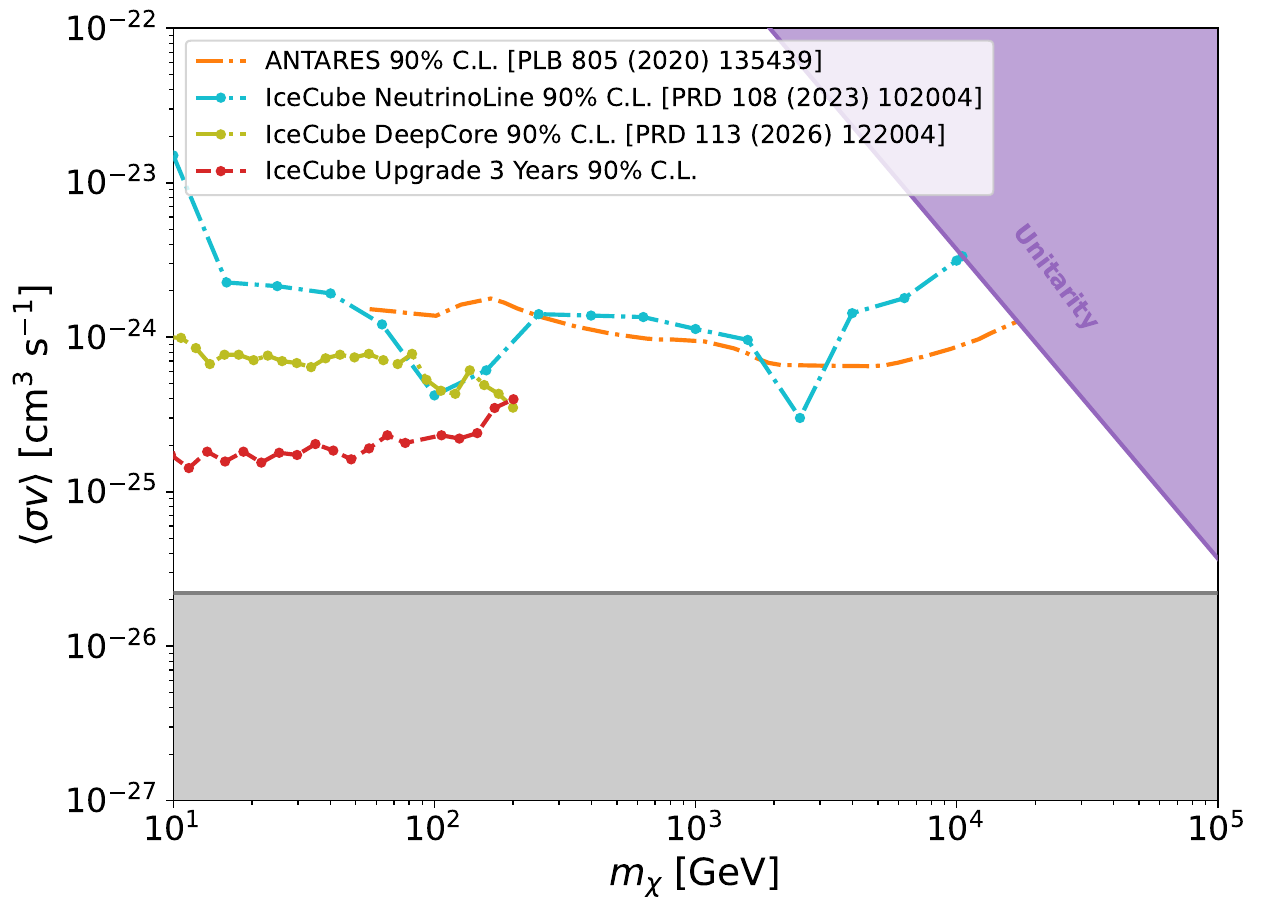}
   \caption{Constraints on $s$-wave annihilation cross section of WIMP DM. The purple shaded region is excluded by unitarity bound~\cite{griest:unitarity}. In the gray shaded region the WIMP is overproduced. \textbf{Left:} Upper bounds for annihilation into visible final states. Blue shaded region (Visibles) is excluded by combining data from Planck measurements of the CMB, {\it Fermi}-LAT observations of dSphs, and AMS-02 positron measurements, taken from ~\cite{Leane:2018kjk}.  Limits from the $\tau^+\tau^-$ annihilation channel are indicated by the pink line for the updated {\it Fermi}-LAT legacy dwarf analysis~\cite{Circiello:2026inp}, the brown line for the combined observations of dSphs from {\it Fermi}-LAT, H.E.S.S., MAGIC, VERITAS, and HAWC~\cite{Fermi-LAT:2025gei}. The projected sensitivities of CTA~\cite{CTA:2020qlo} and SWGO~\cite{Viana:2019ucn} are shown as dashed red and green lines, respectively. \textbf{Right:} Upper bounds for annihilation into invisible final states.
   Limits from searches for DM annihilation from the Galactic Center in the $\nu_e\bar{\nu}_e$ channel are displayed as the dash-dotted orange line for ANTARES~\cite{ANTARES:2019svn}, the dash-dotted cyan and olive lines for IceCube~\cite{IceCube:2025fcn,IceCube:2023ies}. The red dashed line shows the projected sensitivity of the IceCube Upgrade to DM annihilation in the Galactic Center~\cite{IceCube:2026rbh}. All results in the right panel assume an NFW halo profile.} 
    \label{fig:indirect_detection}
\end{figure}

DSph galaxies provide clean gamma-ray targets with little intrinsic astrophysical emission for exploring the low-mass region of the WIMP parameter space. We show in Fig.~\ref{fig:indirect_detection} the upper limits on $\langle \sigma v\rangle$ for the $\tau^+\tau^-$ annihilation channel, which typically provides the strongest bounds among common annihilation channels. The brown line shows the limits obtained by combining observations of dSphs from five gamma-ray telescopes, including {\it Fermi}-LAT, the ground-based imaging atmospheric Cherenkov telescope arrays H.E.S.S., MAGIC, and VERITAS, and the HAWC water Cherenkov detector~\cite{Fermi-LAT:2025gei}. The pink line shows the results of the updated {\it Fermi}-LAT legacy analysis~\cite{Circiello:2026inp}. For consistency, we plot constraints derived using the dSph $J$-factors from Geringer-Sameth et al.~\cite{Geringer-Sameth:2014yza,Geringer-Sameth:2014qqa}, which come from Jeans modeling of stellar-kinematic data with a generalized NFW profile (for dSphs with direct kinematic measurements, otherwise supplemented in Ref.~\cite{Circiello:2026inp} by the scaling relation of Ref.~\cite{Pace:2018tin}).  

For larger WIMP masses, $100\gev \lesssim m_\chi \lesssim 100\tev$, the inner Galaxy has the potential to provide the leading constraints. The projected sensitivities of CTA~\cite{CTA:2020qlo} and SWGO~\cite{Viana:2019ucn} are shown by the dashed red and green lines, respectively in the left panel of Fig.~\ref{fig:indirect_detection}. The sensitivities are calculated assuming an Einasto density profile and for the $\tau^+\tau^-$ annihilation channel (see Ref.~\cite{SWGO:2025taj} for updated sensitivity of SWGO for $b\bar{b}$ channel). 
While not shown in Fig.~\ref{fig:indirect_detection}, we note that H.E.S.S. measurements of the Galactic Center region may also already substantially constrain the parameter space~\cite{HESS:2022ygk,Dutta:2022wdi,Rodd:2024qsi}.  

For annihilation into invisible final states, we plot the current upper limits on $\langle \sigma v\rangle$ from ANTARES~\cite{ANTARES:2019svn} and IceCube~\cite{IceCube:2025fcn,IceCube:2023ies} searches for DM annihilation from the Galactic Center in the $\nu_e\bar{\nu}_e$ channel, shown as the dash-dotted orange, cyan, and olive lines in the right panel of Fig.~\ref{fig:indirect_detection}. The red dashed line shows the projected sensitivity of the IceCube Upgrade for DM annihilation in the Galactic Center~\cite{IceCube:2026rbh}. All results assume an NFW halo profile.

\begin{figure}[t]
    \centering
    \includegraphics[width=0.45\textwidth]{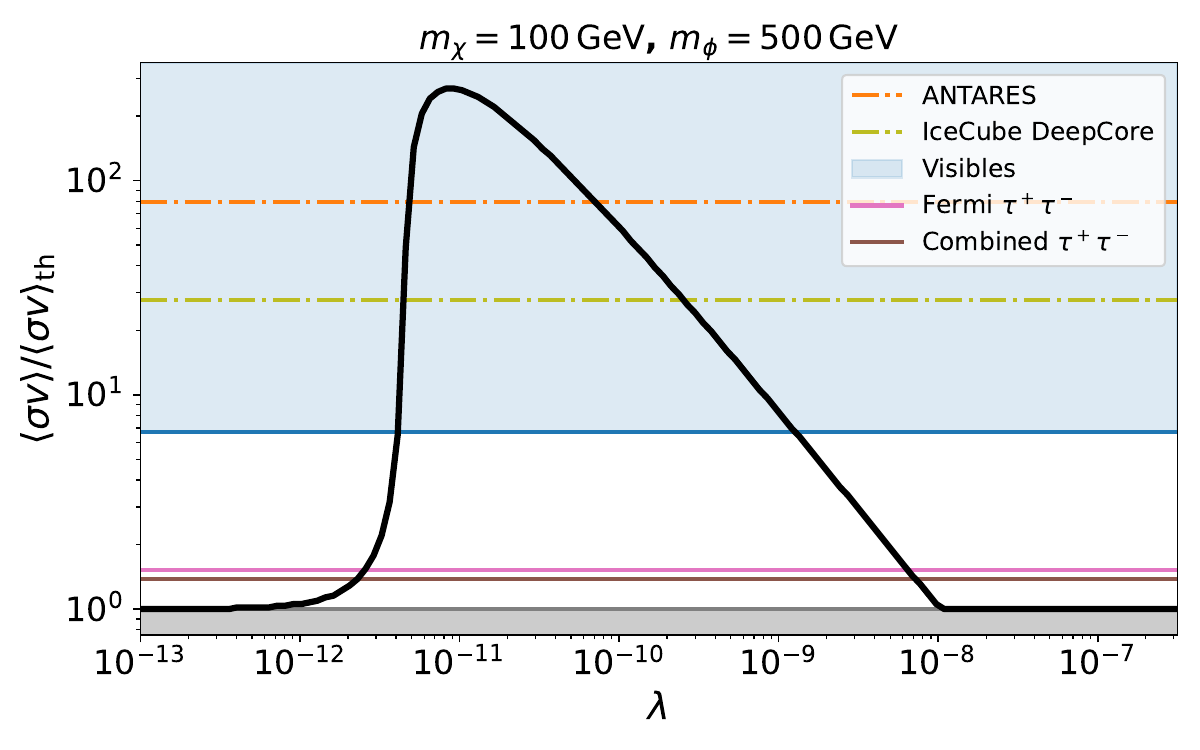}
    \includegraphics[width=0.45\textwidth]{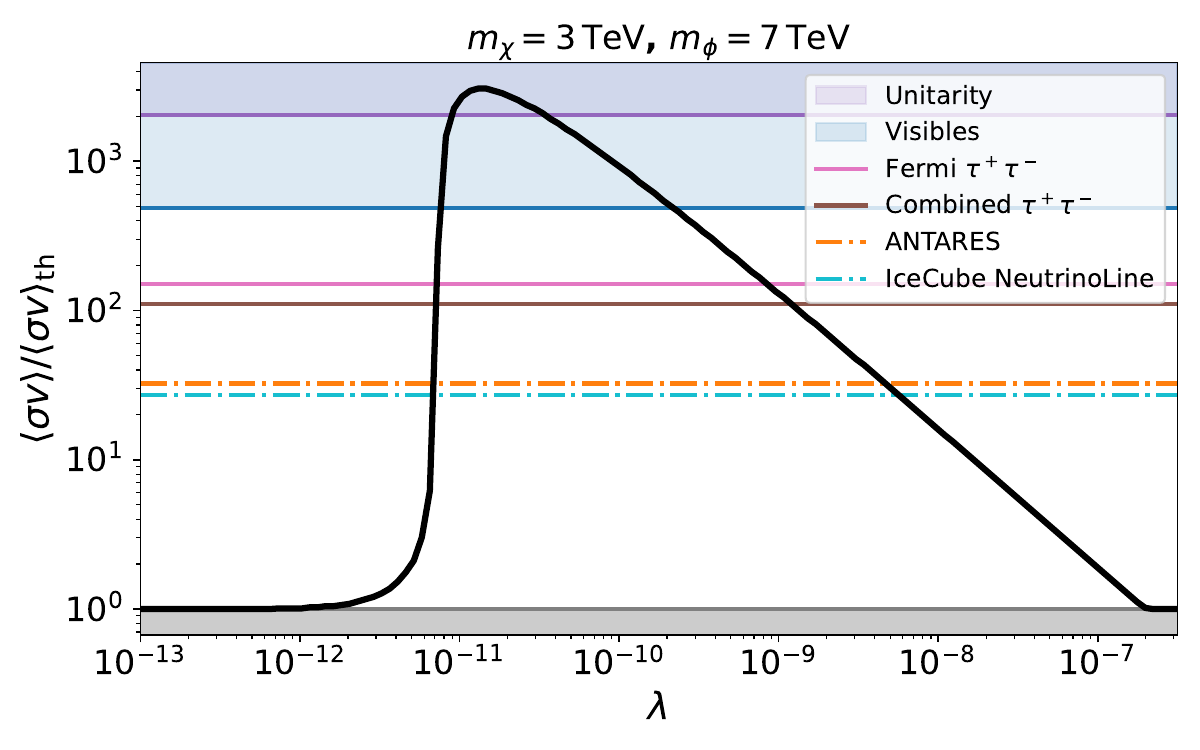}
   \caption{Constraints on the coupling constant between the WIMP and its partner arising from the enhanced annihilation cross section required to yield the observed DM relic abundance. \textbf{Left:}  Corresponds to the benchmark point $\{m_\chi=100~{\rm GeV}, m_\phi=500~{\rm GeV}\}$.  \textbf{Right:} shows the benchmark point $\{m_\chi=3~{\rm TeV}, m_\phi=7~{\rm TeV}\}$. Constraints for each DM mass point are extracted from Fig.~\ref{fig:indirect_detection}.}
    \label{fig:wimp_ext}
\end{figure}

The partner particle of the WIMP in our scenario can be indirectly probed by the DM indirect detection searches discussed above. For fixed WIMP and partner masses, requiring the correct DM abundance today determines the required annihilation cross section as a function of the coupling $\lambda$, allowing us to identify the values of $\lambda$ that are ruled out by the indirect detection constraints. We illustrate this in Fig.~\ref{fig:wimp_ext} by comparing the required WIMP annihilation cross section as a function of the coupling $\lambda$ for two sets of mass parameters with the aforementioned constraints. The left and right panels correspond to the benchmark points $\{m_\chi=100~{\rm GeV}, m_\phi=500~{\rm GeV}\}$  and $\{m_\chi=3~{\rm TeV}, m_\phi=7~{\rm TeV}\}$, respectively. For mass parameters $\{m_\chi=100~{\rm GeV}, m_\phi=500~{\rm GeV}\}$, and annihilation into visible final states, coupling constants in the range $4\times 10^{-12}\lesssim\lambda \lesssim 10^{-9}$ are excluded. This range can be extended to $2\times 10^{-12}\lesssim\lambda \lesssim 7\times 10^{-9}$ for the $\tau^+\tau^-$ annihilation channel. For annihilation into invisible final states, however, the excluded range shrinks to  $4\times 10^{-12}\lesssim\lambda \lesssim 2\times 10^{-10}$. For benchmark mass parameters $\{m_\chi=3~{\rm TeV}, m_\phi=7~{\rm TeV}\}$, the unitarity bound excludes coupling constants in the range $8\times 10^{-12}\lesssim\lambda \lesssim 3\times 10^{-11}$. Assuming annihilation into visible final states, the excluded range can extend to $8\times 10^{-12}\lesssim\lambda \lesssim 2\times 10^{-10}$, and can be further extended for the $\tau^+\tau^-$ annihilation channel to  $7\times 10^{-12}\lesssim\lambda \lesssim 10^{-9}$. For annihilation into invisible final states, the excluded range is $7\times 10^{-12}\lesssim\lambda \lesssim 4\times 10^{-9}$.

Taking constraints from dSph galaxies and the Galactic Center together, current and future data will probe substantial parts of the relevant mass--cross section plane. However, the resulting limits are subject to significant uncertainties in the DM distribution, final-state composition, and astrophysical backgrounds. For dSph galaxies, the inferred DM distribution can be poorly constrained, particularly for ultra-faint systems with small stellar samples~\cite{Bonnivard:2015xpq}. The resulting $J$-factor uncertainties can exceed an order of magnitude for the least-constrained systems~\cite{Pace:2018tin}. Different assumptions about the stellar population, halo profile, and spatial extent of the dSph galaxies can change the resulting limits by factors of a few~\cite{Bonnivard:2015vua,Ichikawa:2016nbi,Ando:2020yyk,DiMauro:2022hue}. The treatment of astrophysical backgrounds introduces additional uncertainties: data-driven approaches can weaken constraints by a factor of a few~\cite{Mazziotta:2012ux, Geringer-Sameth:2014qqa,Boddy:2018qur, Calore_2018, Alvarez_2020, Boddy:2019kuw, Hoskinson:2024hpk}, while background mismodeling can shift limits by up to a factor of two~\cite{Linden:2019soa}. While increased gamma-ray exposure improves statistical sensitivity, it does not by itself eliminate uncertainties associated with stellar membership, halo priors, spatial templates, and diffuse-background modeling, which require improved spectroscopy, dynamical modeling, simulations, and background treatment. Similar uncertainties affect Galactic Center searches, where the inner DM profile is not well determined~\cite{Benito:2020lgu} and different models can lead to substantial variations in the predicted $J$-factor~\cite{Hussein:2025xwm}. Additional degeneracies are due to bright, structured diffuse emission and unresolved point sources where varying the diffuse-emission model can change the inferred excess normalization by factors of a few~\cite{Calore:2014xka}, while foreground mismodeling can mimic or obscure a DM-like component~\cite{Buschmann:2020adf}. Consequently, recent analyses continue to debate the origins of the Galactic Center Excess~\cite{Hooper:2022bec,Manconi:2025ogr,Holst:2024fvb}. Taken together, these uncertainties mean that the limits presented here should be regarded as subject to astrophysical and modeling uncertainties rather than as rigid exclusions.

In this context, it is also interesting to note that our model can provide an interpretation of a recently reported statistically significant, halo-like gamma ray excess in the Milky Way, with the energy spectrum peaking at $20~\gev$~\cite{Totani:2025fxx}. This excess can be fitted by DM annihilation with a DM mass of $m_\chi \sim 0.5$--$0.8\tev$ and an annihilation cross section of $\langle\sigma v\rangle\simeq (5$--$8)\times10^{-25}~\rm cm^3\,s^{-1}$ for the $b\bar{b}$ annihilation channel~\cite{Totani:2025fxx}. We emphasize, however, that this interpretation remains tentative and further investigation is needed to establish whether the reported excess is indeed present and, if so, whether DM annihilation can provide a viable explanation.
\subsection{Cosmological Constraints}
A relatively late decay of an unstable particle such as $\phi$ can modify cosmology and can therefore be constrained by cosmological observations. A long-lived unstable particle might dominate the energy density of the Universe after becoming non-relativistic and prior to its decay. In that case, we would have an early matter-dominated era. To determine under what conditions this can happen in our scenario, we require $x_{\rm eq}<x_{\rm dec.}$, where $x_{\rm eq}$ marks the time at which the energy density of $\phi$ particles becomes equal to the energy density of radiation. It can be found by solving $\rho_\phi(x)=\rho_{\rm rad.}(x)$, where $\rho_\phi(x)=m_\phi Y^{\rm inv.}_\phi(\infty)s(x)$ and $\rho_{\rm rad.}(x)=(\pi^2/30)g_\star m_\chi^4/x^4$ are energy densities of $\phi$ and radiation, respectively. Requiring matter domination prior to decay is equivalent to imposing a lower bound on the coupling, $\lambda\geq\lambda_{\rm MD}$. However, it is straightforward to show that such a coupling is sufficiently large to bring the partner particle into equilibrium with the bath. Therefore, we conclude that there is no consistent scenario in which $\phi$ is both long-lived and dominates the energy density of the Universe before its decay.

If $\phi$ decays around the time of Big Bang Nucleosynthesis (BBN), when the temperature of the Universe is $T=T_{\rm BBN}\sim\rm{MeV}$, it can affect the successful predictions of BBN. Since $\phi$ decays exclusively into WIMPs, its decay is only expected to impact BBN if it triggers a re-annihilation phase around the BBN epoch; in contrast, the freeze-in scenario is essentially harmless. To determine the range of parameters for which BBN constraints become relevant, we first require the decay to occur no earlier than the onset of BBN,
$x_{\rm dec}\gtrsim x_{\rm BBN}\equiv m_\chi/T_{\rm BBN}$, which places an upper bound on the coupling, denoted by $\lambda_{\rm BBN}$. Using Eq.~(\ref{eq:lambda_threshold}) and requiring this bound to lie in the re-annihilation regime, $\lambda_{\rm BBN}> \lambda^{\rm min}_{\rm re\text{-}ann.}$, then implies $m_\chi \lesssim 1.6\,\rm{GeV}$, which points to the sub-GeV mass range. Therefore, for the GeV-TeV WIMP masses considered in this work, BBN does not impose a relevant constraint.

The injection of ionizing particles from DM annihilation during the cosmic dark ages modifies the ionization history and, consequently, perturbs CMB anisotropies. Therefore, measurements of CMB anisotropies provide robust constraints on the production of ionizing particles from DM annihilation. These constraints are included among the visible constraints (Visibles) discussed and imposed at the beginning of this section.

\section{Connection to Baryogenesis}
\label{sec:baryogenesis}
Although the origin of DM and that of ordinary matter, i.e., baryogenesis, do not necessarily have to be connected, the similarity in their observed abundances motivates models that relate the two. One interesting possibility is connecting baryogenesis to WIMP DM, so-called WIMPy baryogenesis~\cite{Cui:2011ab,Choi:2018kto}. In such scenarios, the annihilation of WIMPs out of thermal equilibrium during freeze-out, if accompanied by baryon number violation and $C$ and $CP$ violation, can satisfy the Sakharov conditions~\cite{Sakharov:1967dj} and potentially explain the baryon asymmetry of the Universe. 

In WIMPy baryogenesis~\cite{Cui:2011ab}, as WIMPs approach freeze-out and depart from thermal equilibrium, their annihilations can generate a baryon asymmetry via
\begin{equation}
    \chi\chi\rightarrow B\psi,
\end{equation}
where $B$ is a baryon-carrying SM particle (quark), and $\psi$ is an exotic, new baryon, followed by possible decays of the exotic
baryon (either through baryon-preserving or baryon-violating interactions). Although WIMP annihilation can produce a baryon asymmetry, there are other processes that have the
potential to wash out the asymmetry, and their freeze-out is crucial to create the observed baryon asymmetry. The two leading sources of washout are inverse annihilations of baryons into WIMPs,
\begin{equation}
   B\psi\rightarrow\chi\chi,
\end{equation}
and baryon-to-antibaryon scattering,
\begin{equation}
   B\psi\rightarrow {\bar B} {\bar \psi}.
\end{equation}
Washout processes must be suppressed to generate a sizeable baryon asymmetry. Once washout processes freeze out, the baryon asymmetry produced by subsequent DM annihilations is no longer erased, and its final value depends on the DM abundance at the time of washout freeze-out. Therefore, washout freeze-out must occur before WIMP freeze-out to ensure that a large baryon asymmetry accumulates.

Early freeze-out of the washout processes can arise from kinematics. If the particle $\psi$ is heavier than the WIMP, its number density becomes Boltzmann suppressed once the temperature drops below $m_\psi$, leading to a suppression of the washout rate. On the other hand, $\psi$ cannot be too heavy, since DM annihilation into $\psi$ particles is kinematically forbidden for $m_\psi \gtrsim 2m_\chi$. Therefore, efficient WIMPy baryogenesis points to the mass range $m_\chi \lesssim m_\psi \lesssim 2m_\chi$. It is therefore expected that, for the mass range $m_\psi \ll m_\chi$, the system is in the strong washout regime and the generated baryon asymmetry is highly suppressed. This is because, although WIMPs annihilate out of equilibrium, $\psi$ remains in thermal equilibrium with the bath, allowing the washout processes to remain highly efficient.

In our scenario, re-annihilation of WIMPs also constitutes an out-of-equilibrium process. However, unlike in the standard WIMPy baryogenesis scenario, re-annihilation occurs at much later times. As a result, even if the exotic state $\psi$ is significantly lighter than the WIMP, the dominant washout processes may have already frozen out by the time re-annihilation takes place. In the conventional WIMPy baryogenesis scenario, a light $\psi$ ($m_\psi \ll m_\chi$) remains in thermal equilibrium for an extended period, keeping the washout processes efficient while WIMPs annihilate and thereby suppressing the generated baryon asymmetry. In contrast, delayed freeze-out from re-annihilation in our framework allows most washout processes to become ineffective before WIMP freeze-out from re-annihilation occurs. Consequently, a sufficient baryon asymmetry may be generated even for $m_\psi \ll m_\chi$. We leave a thorough exploration of baryogenesis in the context of our scenario and its implications for future work.
\section{Particle Physics Model}
\label{sec:NMSSM}
As a realization of our scenario, we focus on supersymmetric (SUSY) models, since they provide well-motivated and natural WIMP DM candidates. In particular, within the Minimal Supersymmetric Standard Model (MSSM), the lightest supersymmetric particle (LSP), typically a neutralino, is a natural DM candidate. This has motivated a broad experimental program, including collider searches as well as direct and indirect detection experiments. The absence of a confirmed signal has significantly constrained large regions of the WIMP and MSSM parameter spaces. 

Our scenario requires a stable weak-scale particle with weak interactions with the SM (WIMP), as well as a new heavy scalar that couples feebly to the WIMP and only negligibly to the SM (either not at all or much more weakly than to the WIMP). Such a setup cannot be embedded within the MSSM, since the MSSM does not contain a scalar with these properties. The next-to-MSSM (NMSSM)~\cite{Ellwanger:2009dp,Maniatis:2009re}, on the other hand, provides all the ingredients required to realize our scenario.

The NMSSM is motivated by the so-called $\mu$-problem of the MSSM. The MSSM requires two Higgs $SU(2)$-doublets, $H_u$ and $H_d$, the vacuum expectation values (vevs) of which generate masses for up-type quarks and down-type quarks
and charged leptons, respectively. The MSSM superpotential requires a mass term $\mu H_uH_d$ to give masses to
up- and down-type fermions. For successful electroweak symmetry breaking, the parameter $\mu$ must be of the order of the soft SUSY-breaking scale. If $\mu$ is much larger, its contribution to the Higgs mass parameters overwhelms the negative soft SUSY-breaking mass-squared terms, preventing the Higgs potential from developing the desired electroweak symmetry-breaking minimum. This issue that $\mu$ has to take a value at the electroweak scale, before spontaneous symmetry breaking occurs is the origin of $\mu$-problem. In the NMSSM, the $\mu$-term is generated dynamically by the vev of a singlet scalar field, $S$ which is the complex scalar component of a gauge singlet chiral superfield $\hat S$.  The superpotential of the $\mathbb{Z}_3$-invariant NMSSM is related to the superpotential of MSSM by~~\cite{Ellwanger:2009dp,Maniatis:2009re}
\begin{equation}
    W_{\rm NMSSM}= W_{\rm MSSM}+\lambda\hat S \hat H_u\cdot\hat H_d+\frac{\kappa}{3}\hat S^3,
    \label{eq:superpotential}
\end{equation}
where $\,\hat \space\,$ denotes a superfield. The soft SUSY breaking terms are given by
\begin{equation}
     V_{\rm soft}=m^2_{H_u}|H_u|^2+m^2_{H_d}|H_d|^2+m^2_{S}|S|^2+(\lambda A_\lambda H_u\cdot H_d S +\frac{1}{3}\kappa A_\kappa S^3+{\rm h.c.}).
\end{equation}
A vev of $S$, denoted by $s$,  generates an effective $\mu$-term,
\begin{equation}
    \mu=\lambda s.
\end{equation}
To determine which region of the NMSSM parameter space can be mapped onto our scenario, in the rest of this section we outline the necessary conditions and their possible implications, while leaving a thorough investigation for future work. We require a scalar field that interacts feebly with a WIMP DM candidate. In the NMSSM, the singlet scalar can play the role of the feebly coupled partner and therefore we are interested in the limit $\lambda\ll1$\footnote{The coupling $\lambda$ in Eq.~(\ref{eq:superpotential}) should not be confused with the coupling constant in our model. To avoid this confusion, we denote the corresponding coupling between the partner and the WIMP in the NMSSM by $g$ with an appropriate subscript.}. We therefore identify the conditions under which the singlet scalar has a feeble interaction with the LSP, while its couplings and mixing with the other scalars, which would allow it to decay into SM particles, remain subdominant compared with its interaction with the LSP. We begin by considering the Higgs potential of the NMSSM,  Higgs mass matrices, mixing between scalar fields and their couplings. We then turn to the neutralino sector of NMSSM and LSP candidates.
 
After expanding the full potential around the real vevs of the neutral physical Higgs fields $v_u, v_d, s$, via $H^0_u=v_u+(H_{uR}+iH_{uI})/\sqrt{2}$, $H^0_d=v_d+(H_{dR}+iH_{dI})/\sqrt{2}$, and $S=s+(S_R+iS_I)/\sqrt{2}$ where index $R$ ($I$) indicates the CP-even (odd) state, one can find the  CP-even mass matrix in the basis $(H_{dR}, H_{uR}, S_R)$. The SM Higgs direction is given by $H_{\rm SM}=H_{dR} \cos\beta +H_{uR}\sin \beta$ where $\tan \beta=v_u/v_d$. In the Higgs basis $(H_{\rm SM},H_{\rm NSM},S_R)$ the singlet scalar $S_R$ mixes with the SM Higgs, $H_{\rm SM}$, via a mass term $\mathcal{M}^2_{H_{\rm SM}S_R}$,  
\begin{equation}
    \mathcal{M}^2_{H_{\rm SM}S_R} =\cos\beta \mathcal{M}^2_{H_{dR}S_R}+\sin\beta\mathcal{M}^2_{H_{uR}S_R}=\frac{\lambda v}{\sqrt{2}}\left[2\mu-\left(A_\lambda+2\kappa s\right)\sin 2\beta\right]. 
\end{equation}
The resultant mixing angle is given by:
\begin{equation}
    \tan2\theta_{H_{\rm SM}S_R}=\frac{2\mathcal{M}^2_{H_{\rm SM}S_R}}{m^2_{S_R}-m^2_{H_{\rm SM}}},
\end{equation}
which for small mixing angle, $|\theta_{H_{\rm SM}S_R}|\ll1$, reduces to
\begin{equation}
    \theta_{H_{\rm SM}S_R}\simeq \frac{\mathcal{M}^2_{H_{\rm SM}S_R}}{m^2_{S_R}-m^2_{H_{\rm SM}}}\propto\lambda.
\end{equation}
This mixing opens up decay channels for the singlet scalar into SM particles through the SM Higgs with $\Gamma_{S_R\rightarrow {\rm SM}}\simeq \theta^2_{H_{\rm SM}S_R}\Gamma_{{H_{\rm SM}\rightarrow {\rm SM}}}(m_{S_R})$. Since the mixing angle is proportional to $\lambda$, even for small values of $\lambda$, the branching ratio of the mixing-induced decays can be comparable to that of decays into LSPs. Therefore, to block these decay channels, we impose the alignment condition $\mathcal{M}^2_{H_{\rm SM}S_R}=0$ or equivalently 
\begin{equation}
    A_\lambda+2\kappa s=\frac{2\mu}{\sin2\beta}.
\end{equation}
The NMSSM potential also introduces a coupling between the singlet and the SM Higgs such that if the singlet is heavier than the two times the mass of the SM Higgs, it will decay. One can show that the coupling between $H_{\rm SM}$ and $S_R$ is also set by the mass mixing parameter,
\begin{equation}
    g_{S_R H_{\rm SM} H_{\rm SM}}=\frac{\mathcal{M}^2_{H_{\rm SM}S_R}}{v}.
\end{equation}
Although the non-SM doublet has no couplings to gauge bosons in exact alignment, it still has ordinary Yukawa couplings to quarks and leptons. Mixing between $S_R$ and $H_{\rm NSM}$ can therefore lead to decay of $S_R$ into SM fermions. The mass term responsible for this mixing is given by
\begin{equation}
    \mathcal{M}^2_{H_{\rm NSM}S_R} =-\frac{\lambda v}{\sqrt{2}}\left(A_\lambda+2\kappa s\right)\cos 2\beta,
\end{equation}
which results in the following mixing angle:
\begin{equation}
        \theta_{H_{\rm NSM}S_R}\simeq \frac{\mathcal{M}^2_{H_{\rm NSM}S_R}}{m^2_{S_R}-m^2_{H_{\rm NSM}}}.
\end{equation}
After imposing the alignment condition, we obtain
\begin{equation}
    \mathcal{M}^2_{H_{\rm NSM}S_R} =-\sqrt{2}\lambda v\mu\cot 2\beta,
\end{equation}
and consequently,
\begin{equation}
        \theta_{H_{\rm NSM}S_R}\simeq \frac{-\sqrt{2}\lambda v\mu\cot 2\beta}{m^2_{S_R}-m^2_{H_{\rm NSM}}},
\end{equation}
which is generally nonzero. The mixing is suppressed when $\sqrt{2}\lambda v\mu\cot 2\beta\ll |m^2_{S_R}-m^2_{H_{\rm NSM}}|$. Choosing an appropriate hierarchy among the scalar masses can also close the decay channels of $S_R$ to other Higgs states.

The neutralino sector of NMSSM includes a singlino, $\tilde S$, and therefore the neutralino in the gauge eigenstate basis can be considered as $\psi^0=(\tilde B, \tilde W^3, \tilde H_d, \tilde H_u, \tilde S)$. The symmetric neutralino mass matrix is given by
\[
\begin{pmatrix}
  M_1 & 0 & -g_1v_d/\sqrt{2} &g_1v_u/\sqrt{2} & 0 \\
   & M_2 & g_2v_d/\sqrt{2} &-g_2v_u/\sqrt{2} & 0  \\
   &  & 0 & -\mu & -\lambda v_u\\
   &  &  & 0 & -\lambda v_d\\
    &  &  &  & 2\kappa s\\
\end{pmatrix},
\]
where $M_1$ and $M_2$ are gaugino soft-breaking masses. A rotation matrix $N$ can be introduced to diagonalize the mass matrix, and the resulting mass eigenstates are given by
\begin{equation}
    \tilde \chi^0_i=N_{ij}\psi^0_i,
\end{equation}
with $\tilde \chi^0_1$ being the lightest neutralino. 

After rotating the neutralino interaction eigenstates into mass eigenstates, we find that the coupling of interactions between the singlet-like CP-even scalar, $S_R$, and the LSP is
\begin{equation}
    g_{S_R\tilde \chi^0_1\tilde \chi^0_1}\sim \lambda N_{13}N_{14}-\kappa N^2_{15}.
\end{equation}
We now outline the conditions under which our scenario could be realized within the NMSSM for different LSP compositions.

{\it\textbf{Higgsino-like LSP.}}
For a mostly higgsino LSP, we have:
\begin{equation}
    |N_{13}|^2+|N_{14}|^2\simeq 1,\quad |N_{11}|^2, |N_{12}|^2, |N_{15}|^2\simeq 0.
\end{equation}
Therefore, if the LSP is mostly higgsino, the coupling between it and its scalar partner is $g_{S_R\tilde \chi^0_1\tilde \chi^0_1}\sim \lambda N_{13}N_{14}\sim \lambda$ (the coupling $g_{S_R\tilde \chi^0_1}$ corresponds to the Yukawa coupling in our Lagrangian). So, the NMSSM with $\lambda\lesssim 10^{-8}$ (for a higgsino with a mass of $100\,{\rm GeV}$) points toward our scenario. To explain the $\mu$-problem successfully, we need a weak-scale effective $\mu$, which leads to a large vev; $s\gtrsim 10^7\,{\rm TeV}$\footnote{In principle, the singlet may acquire an initial displacement from its late-time vacuum (due to a slight mismatch between the inflationary minimum and the late NMSSM minimum), which would contribute to its initial abundance. In this study, we assume that the abundance of the scalar partner is initially negligible and builds up gradually through freeze-in from the WIMPs. An initial condensate of the scalar can contribute to its final abundance and, while it may be harmless and allow for smaller values of the coupling to the WIMP, it could also decay at late times or come to dominate the Universe. The phenomenology associated with an initial displacement of the singlet scalar from its vacuum is interesting and will be left for future work.}. To ensure that the LSP is mostly higgsino ($m_{\tilde \chi^0_1}\simeq \mu$), we require $\mu<M_1,M_2, 2\kappa s$ where $m_{\tilde{S}}\simeq2\kappa s$. This leads to the requirement
\begin{equation}
    \frac{\kappa}{\lambda}>\frac{1}{2}.
\end{equation}

Finally, we need to make sure that $S_R$ can decay into LSPs: $m_{S_R}\geq2 m_{\tilde \chi^0_1}\simeq 2\mu$. For the mass of the scalar, we have
\begin{equation}
    m_{S_R}\simeq \sqrt{A_\lambda \lambda\frac{v_uv_d}{s}+\kappa s(A_\kappa+4\kappa s)}\simeq \sqrt{\kappa s(A_\kappa+4\kappa s)}.
\end{equation}
Therefore, decay into higgsinos leads to the condition
\begin{equation}
    \sqrt{\kappa s(A_\kappa+4\kappa s)}\geq 2\mu.
\end{equation}
When this condition is combined with the requirement of no tachyonic pseudoscalars and the absence of the decay channel $S_R\rightarrow 2S_I$, one obtains the stronger constraint $\kappa/\lambda>1$.

{\it\textbf{Bino-like LSP.}}
The NMSSM singlet has no renormalizable coupling to binos. Therefore, a pure bino LSP cannot serve as our WIMP candidate. However, a mostly bino-like LSP, described by 
\begin{equation}
    |N_{11}|^2\simeq 1,\quad |N_{12}|^2, |N_{13}|^2, |N_{14}|^2, |N_{15}|^2\simeq 0.
\end{equation}
with a small higgsino admixture can realize our scenario. In this case, 
\begin{equation}
    g_{S_R\tilde \chi^0_1\tilde \chi^0_1}\sim \lambda N_{13}N_{14}.
\end{equation}
We require $M_1<\mu,M_2, 2\kappa s$ and $m_{S_R}>2M_1$. Also, the direct higgsino channel should be blocked: $m_{S_R}<2\mu$. Since the bino realization requires neutralino mixing, it is less straightforward than the higgsino case, where the coupling between the WIMP and the scalar is directly identified with the NMSSM parameter $\lambda$.

{\it\textbf{Singlino-like LSP.}} 
For a mostly-singlino LSP, we have
\begin{equation}
    |N_{15}|^2\simeq 1,\quad |N_{11}|^2, |N_{12}|^2, |N_{13}|^2, |N_{14}|^2\simeq 0.
\end{equation}
Therefore the coupling between the singlet scalar and the WIMP candidate is given by
\begin{equation}
    g_{S_R\tilde \chi^0_1\tilde \chi^0_1}\sim-\kappa N^2_{15}\sim -\kappa.
\end{equation}
Requiring a feeble interaction between the scalar and the WIMP candidate implies that $\kappa$ must be small. For large values of $\lambda$, the singlino mass is much smaller than the weak scale, while the interaction between the singlet scalar and the higgsino becomes stronger. On the other hand, for small $\lambda$, with $\lambda\sim\kappa$, the singlino is typically heavier than the higgsino and cannot efficiently thermalize with the bath. Moreover, the singlet scalar is not sufficiently heavy to decay into singlinos.

These considerations lead us to conclude that our scenario could potentially be realized in the context of the NMSSM with a mostly higgsino LSP, for a small coupling $\lambda$ and a $\kappa$ parameter larger than $\lambda$. To explain the $\mu$-problem, the singlet has to acquire a large vev. It is worth noting that the decoupling limit, defined by $\lambda\to 0$, $\kappa\to 0$, and $s\to\infty$, in which the NMSSM effectively reduces to the MSSM, also provides a realization of our scenario, provided that $\kappa/\lambda\sim\mathcal{O}(1)$. 

We now discuss the implications of realizing our scenario within the NMSSM. Since the annihilation cross section of the LSP is set by its weak-scale interactions, it cannot be increased arbitrarily. If a region of the NMSSM parameter space allows the singlet scalar to be gradually populated through interactions with the LSP and subsequently decay back into LSPs, then the resulting non-thermal LSP production can provide an additional contribution to the DM abundance. In particular, if the LSP is underproduced through standard freeze-out from equilibrium, the additional LSPs produced through scalar decays can bring its abundance up to the observed value. On the other hand, if the LSP abundance from standard freeze-out alone is already sufficient to account for the observed DM abundance, additional LSP production from scalar decays would lead to overproduction. Such regions of the NMSSM parameter space are therefore not allowed, as they would overclose the Universe.

In the NMSSM, as in the MSSM, a higgsino-like LSP is generally not an isolated state. Instead, it belongs to an approximately degenerate electroweakino multiplet consisting of the two lightest neutralinos and the lightest chargino. This degeneracy originates from the supersymmetric higgsino mass parameter \(\mu\), while mixing with the gauginos and singlino, together with electroweak radiative corrections, generates small mass splittings among these states. When the heavier higgsino-like states are only slightly heavier than the LSP, their abundances remain comparable to that of the LSP around freeze-out. Consequently, the standard treatment based only on LSP pair annihilation is insufficient, and annihilation processes involving the nearly degenerate neutralino and chargino states, known as coannihilations, must also be included~\cite{Griest:1990kh}. These coannihilation processes significantly increase the effective annihilation cross section and reduce the thermal relic abundance. As a result, in a standard thermal history, a nearly pure higgsino reproduces the observed DM relic abundance for a mass of approximately $m_\chi\simeq 1.1\,{\rm TeV}$~\cite{Mizuta:1992qp,Roszkowski:2017nbc,Kowalska:2018toh}, whereas a sub-TeV higgsino is thermally underproduced and can account for only a fraction of the observed DM. 
In the regime where the NMSSM satisfies the conditions of our scenario, the singlet scalar decays into higgsinos, and the combination of the resulting non-thermal higgsino population and the enhanced effective annihilation cross section can potentially yield the correct relic abundance in a radiation-dominated Universe, even for sub-TeV higgsinos. We leave a detailed investigation of this interesting possibility, including the corresponding region of NMSSM parameter space that accommodates sub-TeV higgsino DM with the correct relic abundance, to future work.
\section{Conclusions}
\label{sec:conclusion}
In this work, we have extended the WIMP paradigm by introducing a new unstable particle that interacts exclusively with WIMPs and is sufficiently heavy to decay into them. We consider a standard cosmology with a reheating temperature well above the mass of the partner particle and assume a negligible initial abundance of the partner. WIMPs, which are initially in thermal equilibrium with the bath, then serve as a portal for producing the partner particle through inverse decays and gradually populating its abundance. The partner particle eventually decays back into WIMPs. Although the partner itself does not play a significant role in cosmological evolution, its decay into WIMPs can substantially alter their thermal history and final abundance.
 
When the partner decays after the WIMPs have frozen out from equilibrium, non-thermal WIMPs injected through its decay may, depending on the lifetime of the partner, initiate a secondary annihilation phase, the so-called re-annihilation. During re-annihilation, the DM yield follows a quasi-equilibrium value and eventually freezes out from it, setting the final DM abundance. If the partner's lifetime is too long, re-annihilation does not occur. Instead, the WIMPs produced through its decay simply freeze in and add to the DM abundance resulting from the initial freeze-out. As we have shown, in either case, re-annihilation or freeze-in, the non-thermally produced WIMPs lead to an overproduction of DM. Therefore, a larger annihilation cross section than the canonical thermal cross section is required to reproduce the observed DM abundance today. The required enhancement, which can be as large as three orders of magnitude, is subject to both theoretical and observational constraints. Ongoing and upcoming gamma-ray and neutrino telescopes can probe the partner particle and constrain the parameter space of this scenario.

Furthermore, we showed that a class of particle physics models that attempt to explain the baryon asymmetry of the Universe and the similar abundances of baryonic matter and DM using WIMP freeze-out can be realized within our framework, with the possibility of extending the parameter space beyond the existing one. Finally, as a realization of our scenario, we considered the NMSSM and showed that, under reasonable assumptions about the model parameters, the singlet scalar together with a higgsino-like LSP as the WIMP candidate can be mapped onto our model. Moreover, we argued that the overproduction of higgsinos due to late-time decays of the NMSSM singlet scalar, combined with coannihilation between higgsinos and nearly degenerate neutralino and chargino states, can accommodate sub-TeV higgsino DM with the correct relic abundance.

Our simple model provides a minimal extension of the WIMP DM paradigm while exhibiting a rich cosmological history and phenomenology, with promising prospects for tests through current and upcoming indirect detection searches. This framework also opens up several interesting directions for future investigation, including a detailed study of its realization in the NMSSM and its implications for sub-TeV higgsino DM, its potential connection to baryogenesis, and its generalization to scenarios with multiple feebly interacting partners of the WIMP.

\acknowledgments
P.S. and C.H. are supported in part by  National Science Foundation grant PHY-2412834. 
B.S.E. is particularly grateful to Pearl Sandick for her continuous support and encouragement. 

\appendix
\section{Analytical Calculation of Freeze-out from Re-annihilation}
\label{appx:reanni}
We introduce $\Delta\equiv Y_\chi-Y_{\chi,{\rm QE}}$ which shows the deviation of the yield of DM from the quasi-equilibrium value. Then we have:
\begin{equation}
   \frac{d \Delta}{dx}+\frac{dY_{\chi,{\rm QE}}}{dx}=-\alpha x^{-2}\Delta(2Y_{\chi,{\rm QE}}+\Delta).
   \label{eq:dDeltadx}
\end{equation}
Around the decay time of $\phi$, $Y_\chi$ follows $Y_{\chi,{\rm QE}}$ closely, therefore $\Delta$ and $|d\Delta/dx|$ are small. The solution can be approximated by imposing $d\Delta/dx\simeq 0$ to obtain:
\begin{equation}
      \Delta\simeq-\frac{dY_{\chi,{\rm QE}}/dx}{\alpha x^{-2}(2Y_{\chi,{\rm QE}}+\Delta)}.
      \label{eq:Deltaearly}
\end{equation}
At later times, after freeze-out, $\Delta\simeq Y_\chi\gg Y_{\chi,{\rm QE}}$. Therefore after neglecting $Y_{\chi,{\rm QE}}$ and $dY_{\chi,{\rm QE}}/dx$, we have:
\begin{equation}
    \frac{d \Delta}{dx}\simeq-\alpha x^{-2}\Delta^2.
    \label{eq:dDeltalate}
\end{equation}
The final yield of DM can be found by integrating Eq.~(\ref{eq:dDeltalate}) from $x=x_{f,{\rm re\text{-}ann.}}$ to $x=\infty$ as:
\begin{equation}
   Y_\infty=\frac{x_{f,{\rm re\text{-}ann.}}}{\alpha}.
\end{equation}
To determine $x_{f,{\rm re\text{-}ann.}}$ we use this criterion that at freeze-out $\Delta$ becomes of the order of $Y_{\chi,{\rm QE}}$. In other words, $\Delta(x_{f,{\rm re\text{-}ann.}})= c_2 Y_{\chi,{\rm QE}}(x_{f,{\rm re\text{-}ann.}})$ where $c_2$ is a numerical constant of order unity. By using Eq.~(\ref{eq:Deltaearly}), and applying the freeze-out criterion we get:
\begin{eqnarray}
  \nonumber    \Delta(x_{f,{\rm re\text{-}ann.}})&\simeq&-\frac{\left(dY_{\chi,{\rm QE}}/dx\right)|_{x_{f,{\rm re\text{-}ann.}}}}{\alpha x_f^{-2}\left[2Y_{\chi,{\rm QE}}(x_{f,{\rm re\text{-}ann.}})+c_2Y_{\chi,{\rm QE}}(x_{f,{\rm re\text{-}ann.}})\right]}\\
  \nonumber    &=&-\frac{1}{\alpha (2+c_2)x_{f,{\rm re\text{-}ann.}}^{-2}}\frac{\left(dY_{\chi,{\rm QE}}/dx\right)|_{x_{f,{\rm re\text{-}ann.}}}}{Y_{\chi,{\rm QE}}(x_{f,{\rm re\text{-}ann.}})}\\
    \nonumber  &=&-\frac{1}{\alpha (2+c_2)x_{f,{\rm re\text{-}ann.}}^{-2}}\frac{3}{2x_{f,{\rm re\text{-}ann.}}}\left[1-\frac{\Gamma_{\phi}}{H(x_{f,{\rm re\text{-}ann.}})}\right]\\
      &=&\frac{3b}{2\alpha (2+c_2)x_{f,{\rm re\text{-}ann.}}^{-1}},
      \label{eq:Deltaxf}     
\end{eqnarray}
where the third line is obtained from Eqs.~(\ref{eq:YQE}) and (\ref{eq:Yphi2}), and since the freeze-out happens after the decay of $\phi$, we expect $1-\Gamma_{\phi}/H(x_{f,{\rm re\text{-}ann.}})=-b$ where $b$ is a positive numerical factor. Now, $x_{f,{\rm re\text{-}ann.}}$ can be evaluated as the solution to the equation:
\begin{equation}
    \frac{3b}{2\alpha (2+c_2)x_{f,{\rm re\text{-}ann.}}^{-1}}= c_2 Y_{\chi,{\rm QE}}(x_{f,{\rm re\text{-}ann.}}),
    \label{eq:condition}
\end{equation}
which is given by:
\begin{eqnarray}
 \nonumber   x_{f,{\rm re\text{-}ann.}}&=&\frac{\sqrt{2}\pi^{3/4}}{5^{1/4}\sqrt{3}}g_\star^{1/4}\frac{m_\chi}{\sqrt{m_{\rm Pl}\Gamma_{\phi}}}\sqrt{-W(-z)}\\
  \nonumber  &\simeq& \frac{\sqrt{2}\pi^{3/4}}{5^{1/4}\sqrt{3}}g_\star^{1/4}\frac{m_\chi}{\sqrt{m_{\rm Pl}\Gamma_{\phi}}}\sqrt{-\ln z}\left[1-\frac{\ln(-\ln z)}{2\ln z}\right]\\
    &=& x_{\rm dec.}\sqrt{-\ln z}\left[1-\frac{\ln(-\ln z)}{2\ln z}\right],
\end{eqnarray}
where 
\begin{equation}
    z\equiv \frac{8\pi^{19/2}b^4}{675\sqrt{5}c_2^4(2+c_2)^4}\frac{g^6_{\star,S}}{g_\phi^2g^{7/2}_\star}\frac{m_\phi^4}{m^5_{\rm Pl}\langle\sigma v\rangle^2\Gamma^3_{\phi}},
\end{equation}
and 
$W(\cdot)$ is the Lambert function defined such that $W(x)\exp[W(x)]=x$, and we use the following asymptotic expansion:
\begin{equation}
    W_{-1}(-x)=\ln x-\ln(-\ln x)+...~~~0<x\ll1
\end{equation}
which leads to
\begin{equation}
    \sqrt{W_{-1}(-x)}\simeq \sqrt{-\ln x}\left[1-\frac{\ln(-\ln x)}{2\ln x}\right].
\end{equation}
It is worth mentioning that Eq.~(\ref{eq:condition}) has two solutions, corresponding to the two real branches of the Lambert function, $W_0(\cdot)$ and $W_{-1}(\cdot)$. The freeze-out time corresponds to the larger solution, which is given by the $W_{-1}(\cdot)$ branch. A real solution for $x_{f,{\rm re\text{-}ann.}}$ exists as long as $0<z\leq 1/e$. 

\bibliography{draft}{}
\end{document}